\documentclass[a4paper,fleqn]{cas-sc}
\usepackage[authoryear,longnamesfirst]{natbib}
\usepackage[utf8]{inputenc}
\usepackage{amsfonts}
\usepackage{amssymb}
\usepackage{amsmath}
\usepackage{xcolor}

\usepackage{savesym}
\savesymbol{comment}

\usepackage{algpseudocode}

\usepackage{algorithm}

\newcommand{\myadded}[1]{\textcolor{black}{#1}}

\usepackage{adjustbox}

\def\tsc#1{\csdef{#1}{\textsc{\lowercase{#1}}\xspace}}
\tsc{WGM}
\tsc{QE}

\usepackage{url}
\usepackage{subcaption}

\usepackage{xcolor}
 
\usepackage{placeins}

\begin{document}
\makeatletter
\@namedef{Changes@AuthorColor}{red}
\colorlet{Changes@Color}{red}
\makeatother

\let\WriteBookmarks\relax
\def\floatpagepagefraction{1}
\def\textpagefraction{.001}

\shorttitle{PreBit - A multimodal model with Twitter FinBERT embeddings for extreme price movement prediction of Bitcoin}    

\shortauthors{Zou and Herremans}  

\title [mode = title]{PreBit - A multimodal model with Twitter FinBERT embeddings for extreme price movement prediction of Bitcoin}  




%

\author[1]{Yanzhao Zou}[type=author,
      style=english,
      auid=000,
      bioid=1,
      prefix=,
      orcid=0000-0001-8988-5981,
      ]

\cormark[1]
\cortext[1]{Corresponding author}

\ead{<zouyanzhao@gmail.com>}


\credit{Development, lead author}

\affiliation[a]{organization={Singapore University of Technology and Design},
            addressline={8 Somapah Road}, 
            city={Singapore},
            postcode={487372}, 
            state={Singapore},
            country={Singapore}}

\author[1]{Dorien Herremans}[type=author,
      style=english,
      auid=000,
      bioid=1,
      prefix=,
      orcid=0000-0001-8607-1640,
      twitter=dorienherremans,
      linkedin=dorienherremans,
      ]

\ead{dorien_herremans@sutd.edu.sg}

\ead[url]{dorienherremans.com}

\credit{Advisor, ideation, editing}





\begin{abstract}
Bitcoin, with its ever-growing popularity, has demonstrated extreme price volatility since its origin. \myadded{Extreme price fluctuations have been know to occur due to Tweets from Elon Musk, Michael Saylor, and others. In this paper, we aim to investigate whether we can leverage Twitter data to predict these extreme price movements. Existing social media models often take a shortcut and include sentiment extracted from Tweets. In this work, however, we want to embed the actual Tweets in a domain-informed way, and investigate whether they have an impact.} Hence, we propose a multimodal deep learning model for predicting extreme price fluctuations that takes as input candlestick data, prices of a variety of correlated assets, technical indicators, as well as Twitter content. To train the model, a new dataset of 5,000 tweets per day containing the keyword `Bitcoin' was collected from 2015 to 2021. This dataset, called PreBit, is made available online\footnote{\url{https://www.kaggle.com/datasets/zyz5557585/prebit-multimodal-dataset-for-bitcoin-price}}, as is our model\footnote{\url{https://github.com/AMAAI-Lab/PreBit}}. 
\myadded{Our proposed hybrid multimodal model consists of an SVM model based on price data, which is fused with a text-based Convolutional Neural Network. In the text-based model,} we use the sentence-level FinBERT embeddings, pretrained on financial lexicons, so as to capture the full contents of the tweets and feed it to the model in an understandable way. 
In an ablation study, \myadded{we explore whether adding social media data from the general public on Bitcoin improves the model's ability to predict} extreme price movements. Finally, we propose and backtest a trading strategy based on the predictions of our models with varying prediction threshold and show that it can be used to build a profitable trading strategy with a reduced risk over a `hold' or moving average strategy. 
\end{abstract}


\begin{graphicalabstract}
\includegraphics[width=1\textwidth]{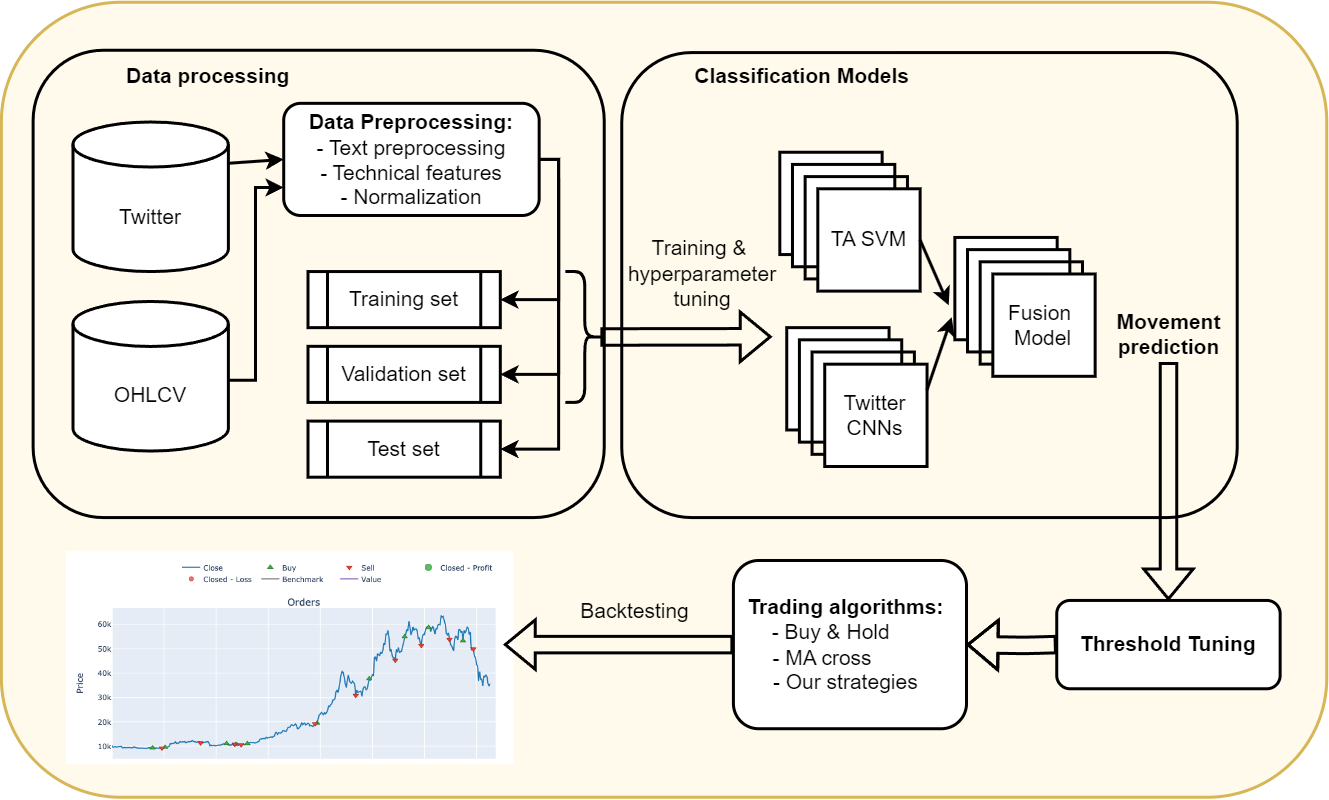}
\end{graphicalabstract}

\begin{highlights}
\item A multimodal model for BTC extreme price movement prediction using Twitter.
\item Ablation study of the impact of different modalities on accuracy.
\item New publicly available dataset of 9,435,437 tweets related to Bitcoin. 
\item A profitable trading strategy with reduced risk exposure for Bitcoin trading. 
\item Demonstrates the influence of predictive thresholds on risk of a trading strategy.
\end{highlights}


\maketitle

\section{Introduction}\label{sec:intro}













\begin{figure}
    \centering
\includegraphics[width=0.8\textwidth]{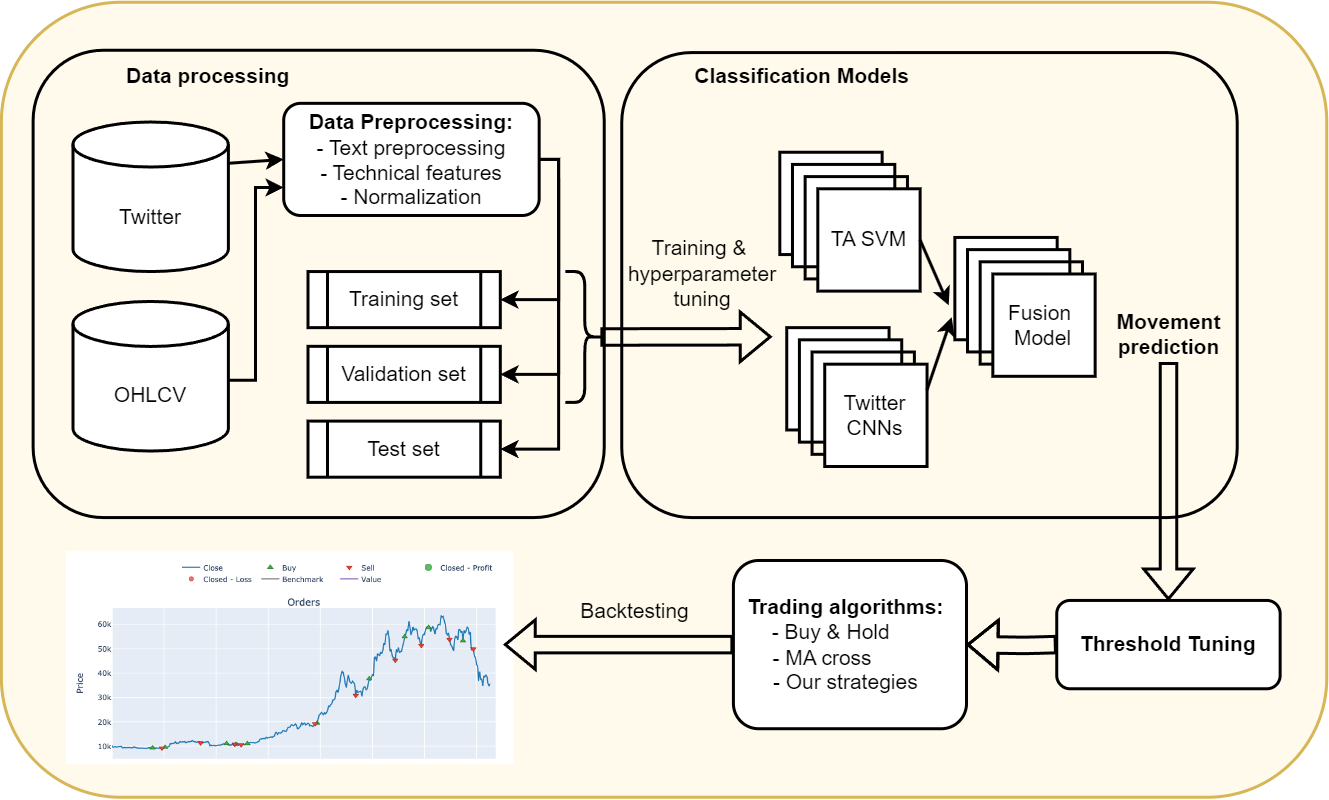}
    \caption{Schematic overview of the paper. }
    \label{fig:overview_diagram}
\end{figure}

With cryptocurrencies gaining traction among both retail and institutional users over the past few years, the market cap of the cryptocurrencies grew significantly. Bitcoin (BTC) is the most traded and largest cryptocurrency by market capitalisation. While trading activities of traditional assets are dominated by institutional investors, retail investors play a much bigger role in Bitcoin trading (see Goldman Sachs report by \citet{goldman21}). Bitcoin is also a digital asset that does not derive its value from physical demands such as coal and iron ore. This makes the Bitcoin price more susceptible to be influenced by the market sentiment. For example,  the price of Bitcoin rose by as much as 5.2 percent on 24 March 2021 when Elon Musk tweeted Tesla would accept Bitcoin for payments. It also crashed as much as 9.5 percent on 13 May 2021 when Elon Musk tweeted to question the energy consumption from Bitcoin mining. In this paper we propose a multimodal model that can predict extreme Bitcoin price movements based on Twitter data as well as an extensive set of price data with technical indicators and related asset prices. 
There exists some research that uses sentiment information from social media to try to predict cryptocurrency prices \citep{mohanty2018predicting}.  
By just using sentiment information, however, a lot of potentially useful information is ignored. In this research, we therefore leverage a state-of-the-art method to embed the \textit{entire tweet contents} into a BERT model (Bidirectional Encoder Representations from Transformers) \citep{devlin2018bert} and use it as input to our predictive model. We further enhance our model using historical candlestick (OHLCV) data and technical indicators, together with correlated asset prices such as Ethereum and Gold. A schematic overview of our work is shown in Figure~\ref{fig:overview_diagram}.

When exploring existing studies that use social media data, we notice that most of the exiting research uses sentiments from \myadded{texts, article titles, or social media posts, or meta-features such as} number of posts, and number of comments as the model input, rather then actual word embeddings. Sentiments are often extracted using pretrained models such as Valence Aware Dictionary for Sentiment Reasoning (VADER) \citep{elbagir2019twitter}, word2vec \citep{acosta2017sentiment}, or BERT \citep{sun2019utilizing}. Firstly, sentiment models pretrained for general purpose may not apply to financial language, for example, they may not accurately model or embed the words `chart', `hold', `bull' or `bear'. \myadded{The distadvantage is that} the context of the text is also lost when only the derived statistics are used. Utilising the full text of posts in the model retains more information and improve model performance. Hence, in this paper, we use the full text embeddings in our predictive model, in combination with a dedicated financial sentence embedding model, FinBERT \citep{araci2019finbert}. An additional challenge when doing this is that the number of words in the tweets gathered every day varies and a neural network typically requires a constant input length. We propose a solution to this problem by concatenating the tweets and splitting them into larger blocks as explained in detail in Section~\ref{sec:model}. To the best of our knowledge, only one study \citep{lamon2017cryptocurrency} has tried to use text embeddings, not just sentiment, to predict Bitcoin prices, \myadded{and this does not use an embedding model pretrained on financial texts}, nor do they predict extreme movements or offer a backtested trading strategy with reduced market exposure risk. This research aims to fill this gap. 

In this paper, we propose a multimodal embedded model for predicting extreme price movements of Bitcoin and evaluate the impact of different modalities, including tweets represented through finBERT context embeddings. 
A new dataset is released consisting of tweets as well as candlestick data, related asset prices (Ethereum and Gold) and a selection of technical indicators from 1 January 2015 until 31 May 2021. In an ablation study, we explore the influence of different multimodal data. The model and dataset is made available online\footnote{\url{https://github.com/AMAAI-Lab/PreBit}}.  In this research, we treat price prediction as a classification problem, whereby we predict next-day extreme price movements (up/down 2 or 5\%), this way, our predictions can be directly embedded in a trading strategy. The proposed (simple) trading strategy  was backtested with different predictive thresholds to optimally control risk exposure. 

The next section provides a review of related literature. In Section~\ref{sec:dataset}, we describe the PreBit dataset in more detail. The proposed models are explained in the next section, followed by our experimental setup in Section~\ref{sec:experiment}. Finally, in Section~\ref{sec:experiments}, the results from the experiments are discussed followed by a conclusion.









\section{Literature Review}\label{sec:lit}


In this section, we will review some of the relevant research on price prediction models and identify how the current research addresses a unique gap. First, we will walk through existing research that uses traditional price information and technical indicators for predicting Bitcoin's price. Then we provide a brief overview of models that use Natural Language Processing (NLP) in traditional stock price prediction models. Lastly, we explore how NLP has been used for cryptocurrency price prediction. 


\subsection{BTC price prediction with technical indicators}

To predict price movements of any asset, a common approach is to look at historical Open-High-Low-Close-Volume (OHLCV) data together with technical analysis (TA) indicators. The same applies to Bitcoin trading. 

Basic machine learning models have been used to construct models that use TA data. For instance, \citet{slepaczuk2018robustness} used six technical features such as Momentum and Relative Strength Index with a Support Vector Machine (SVM) \citep{cortes1995support} model to predict whether several cryptocurrency belonged to the top or bottom quintile in terms of volatility-adjusted returns. However, their model was not able to outperform simple Buy-and-Hold or Equal-Weight strategies. \citet{sun2019using} used OHLCV data to predict price change direction for Bitcoin and Ethereum with a 5-minute frequency, using 16 technical indicators and a Random Forest.  The backtesting results for their strategy demonstrate the capability to capture trend change to a certain extent, but struggles with reversal periods.

With the success of neural networks over the past few years, recent research has also explored the use of deep learning techniques for price prediction. Recurrent neural networks architectures, including Long-Short-Term-Memory (LSTM) \citep{hochreiter1997long}, are a widely adopted model in time-series prediction for their capability to capture temporal information \citep{mikolov2010recurrent}. For predicting digital asset prices, \citet{kwon2019time} used the OHLC and volume data of a basket of cryptocurrencies at a 10-min interval to construct a training dataset and predict price movement direction as a classification problem. The model outperforms a baseline gradient boosting model. Long-Short-Term Memory-Models (LSTMs) are a popular architecture for cryptocurrency price prediction with TAs. \citep{aditya2022time}, \citep{wu2018new} \citep{shin2021bitcoin}, \citep{felizardo2019comparative}.  \citet{alonso2020convolution} combined LSTMs with Convolutional Neutral Networks (CNNs) using OHLC data and 18 derived technical indicators. This architecture outperformed the LSTM and Multi-layer Perceptron (MLP) models for Bitcoin, Litecoin and Ethereum price prediction on a 1-min interval from July 2018 to June 2019. 

New deep learning techniques that capture the temporal relations between data include WaveNet (Temporal CNNs with residual connections) \citep{oord2016wavenet} and Transformer networks \citep{vaswani2017attention} have recently also been used for digital asset prediction. \citet{felizardo2019comparative} did a comparison study between WaveNet, \myadded{Autoregressive Integrated Moving Average} (ARIMA), Random Forest, SVM, and LSTM to predict the future price of Bitcoin using OHLCV data. They found that the ARIMA model and the SVM equally outperformed the other models including WaveNet. One potential reason for this is that deeper models like Wavenet traditionally need a bigger dataset compared to models such as SVMs and ARIMA.  Transformers, however, have demonstrated superior performance in the study conduced by \citet{afteniy2021predicting}. Using only Bitcoin OHLC data on a 1-min interval for 2007 to 2021 as input, the authors reported 83\% accuracy for a Transformer model versus the 67\% accuracy for an LSTM architecture when performing  directional price change prediction. \citet{sridhar2021multi} also used the Transformer architecture with Time2vec on Dogecoin price data and reported a better result in terms of Mean Absolute Error (MAE) compared to an LSTM model \citep{shin2021bitcoin}, \myadded{Recurrent Neural Network} (RNN) model \citep{kavitha2020performance}, and a Linear neural network \citep{ali2020data}. 

\myadded{Despite the advances in time-series prediction, investors typically make use of more information other than just the candlestick data when deciding to buy an asset. Hence, in the next subsection, we will discuss how NLP techniques have been used for predicting asset prices. }



\subsection{Stock price prediction with NLP}

Looking at the stock market, there is a longstanding believe that sentiment and social influence has an impact on prices. For instance, the efficient market hypothesis states that assets are always trading at their fair value, thus leaving outperforming the market impossible \citep{malkiel1989efficient}. Provided an NLP model could capture all the knowledge of the world, we could in theory be able to predict the correct price. There are, however, limitations to this hypothesis: 1) markets may not really be efficient \citep{wang1985some}; 2) we cannot include all the information of the world in our model.



The recent success of the Transformer-based BERT model \citep{devlin2018bert} for multiple NLP tasks has attracted the attention of financial researchers as well. Combining the sentiment information extracted from BERT on carefully annotated Twitter data with stock OHLC price data, \citet{dong2020belt} reported better results in terms of Area Under the Curve (AUC) for next-day price prediction compared to the state-of-the-art stock prediction model StockNet developed by \citep{xu2018stock}. Other similar work using BERT equally reported that their models outperformed others to various degrees~\citep{sonkiya2021stock, chen2021stock}. Based on these results, we have opted to use the state-of-the-art BERT model for this research, pretrained on a financial lexicon. 


In general, we see many sources of text data used in stock prediction research, for instance, StockTwits \citep{jaggi2021text}, Yahoo! Finance news \citep{schumaker2009quantitative}, Reuters and Bloomberg news \citep{ding2015deep}, Twitter data \citep{bollen2011twitter, si2013exploiting,das2018real,oliveira2017impact, gross2019buzzwords, teti2019relationship,valle2021does, pagolu2016sentiment}, Dow Jones Newswire \citep{moniz2014classifying}, and Bloomberg reports \citep{chan2011text}. Given its popularity in financial circuits, we have opted to use Twitter data in this study. 
For a more complete overview of text-based methods for stock prediction, the reader is referred to the survey papers by \citet{de2014evaluating, kumar2016survey, thakkar2021fusion}. \myadded{The next subsection focuses on how some of these techniques have been used for digital assets.}

\subsection{BTC price prediction with \myadded{multimodal models}}

Just like stock prices are influenced by any available multimodal datastreams, so are digital asset prices. \myadded{In addition to the traditional price-based data, we see the emergence of multimodal models for digital assets price prediction, that combine this data with different types of data. Examples of such data include on-chain data as well as exchange data \citep{herremans2022forecasting}, sentiment and social media data such as} Twitter data \citep{mohapatra2019kryptooracle, wolk2020advanced,raju2020real, kim2021predicting}, Telegram \citep{smuts2019drives, patel2020deep}, Reddit\citep{ortu2022technical,raju2020real}, and more. Looking at existing literature, there are a number of models that again use the sentiment score derived from a text source as input to a predictive price model. \myadded{We briefly go over related literature, but for a more comprehensive literature review on cryptocurrency trading research, the reader is referred to \citet{fang2020cryptocurrency}. }

\citet{kim2016predicting} used VADER (Valence Aware Dictionary for Sentiment Reasoning \citep{hutto2014vader}) to generate a sentiment score from comments and replies on the Bitcoin, Ethereum and Ripple community posts. \citet{akbiyik2021ask} also used VADER to to extract twitter sentiment as part of the model inputs in an early fusion set-up for next-fifteen-minute Bitcoin realised volatility prediction. In a more recent study, \citet{aharon2022twitter} examined the relationship between Twitter-based uncertainty and cryptocurrency returns. They used Twitter-Based Economic Uncertainty (TEU) and Twitter-Based Market Uncertainty (TMU), two indices generated by the Economic Policy Uncertainty website from 2011 to 2020 \citep{baker2021twitter}, together with OHLCV data from CoinDesk. The authors documented a significant directional predictability from the TMU and TEU for Bitcoin based on a cross-quantilogram analysis. \citet{nghiem2021detecting} used historical posts from Telegram channels collected using PumpOlymp\footnote{\url{https://pumpolymp.com/}} which included keywords such as  ``pump'', ``dump''and ``signals'' as part of their social media data. The authors built a combined CNN and LSTM model to predict `pumps' with a prediction error typically less than 5\% away from the true price. 
They used statistics of social media data such as number of posts on Twitter, Facebook, Github repositories, number of comments, views and open issues.

\citet{mohanty2018predicting} used sentiments from tweets  with technical indicators in an LSTM that marginally outperforms a random model in predicting the next-ten-minute Bitcoin price change direction. 
\citet{passalis2021learning} used FinBERT~\citep{araci2019finbert} to extract sentiments from a dataset with 223,000 news headlines collected by BDC consulting\footnote{\url{https://bdc.consulting/insights/cryptocurrency/analyzing-crypto-headlines}}. They report a high return for their CNN model price change detection model based on sentiments, which only slightly outperformed the one using OHLCV data. 
 Similarly, \citet{cruz2021financial}'s autoencoder model with sentiment information (from BERT trained on 10,111 news articles) outperforms the model without sentiment information by 3.7 ppt in terms of $R^2$. 
A similar conclusion was made by \citet{ortu2022technical} for there model that uses sentiment from 4,423 Github and 33,000 Reddit user comments. 

\myadded{Using only the sentiment of the available text is popular \citep{ye2022stacking, akbiyik2023ask, haritha2023cryptocurrency, leung2023intelligent, critien2022bitcoin, sabri2022cryptocurrency}, but is very limiting. By using the actual text (content) instead of only the sentiment, much more data could be used to make more accurate predictions. However, we were only able to find one study that} directly uses the actual text and feeds it to a predictive model without first extracting the sentiment.  
\citet{lamon2017cryptocurrency} used a bag-of-words (BoW) to extract embeddings from news headlines from \url{cryptocoin.com} as well as 60 tweets daily on the topic of cryptocurrency. Experimenting with logistic regression, SVM, and Naive Bayes, the authors reported that logistic regression performed the best and was able to consistently achieve higher than 50\% accuracy in predicting next-day Bitcoin price change direction.

\myadded{In the last few years, cutting edge NLP research includes much more effective techniques for embedding text into NLP models, other than bag-of-words. Hence, we turn to some of the latest state-of-the-art in this paper and explore FinBERT embeddings for Twitter. In addition, many of the previous studies do not make their dataset available, so there can be no direct comparison or benchmarking. In this study, our source code and dataset is made available online to allow other researchers to further improve upon our work. }






In the current study, we aim to fully use social media content, beyond just using sentiment scores. We therefore leverage upon the results by \citet{lamon2017cryptocurrency} and improve their approach by using a pretrained BERT on finance data: finBERT, which should be better able to capture financial content \citep{araci2019finbert}. 
Contrary to many other research studies, we also propose a trading strategy based on the models and thoroughly backtest it to illustrate how such models may be used to decrease the downward risk of trading strategies. 


\section{Multimodal Dataset}
\label{sec:dataset}

We present a new dataset, PreBit, which consists of two modalities: daily price, correlated assets with technical analysis data for BTC (which we will refer to as TA data for simplicity), as well as a the contents of a 5,000 of daily tweets. The dataset is available online\footnote{\url{https://www.kaggle.com/datasets/zyz5557585/prebit-multimodal-dataset-for-bitcoin-price}}. In the next two subsections we will discuss these two modalities in more detail.


\subsection{Twitter Data}
\paragraph{Collecting tweets}
The PreBit dataset consists of publicly available tweets containing the keyword `Bitcoin' from 1 January 2015 until 31 May 2021. A total of 5,000 tweets (or the maximum available that day) were collected per day using the GetOldTweets-python library \footnote{\url{https://github.com/Jefferson-Henrique/GetOldTweets-python}}, starting at 23:59:59 GMT+0 and tracking backwards. This resulted in a total of 9,435,437 tweets over the entire period. In 2015 there are a few days with less than 5,000 tweets, which is due to the fact that Bitcoin was not yet as popular in these earlier years. 

Each tweet contains the following attributes: username, timestamp in datetime format (minutes), number of retweets and favourites, tweet content, mentions (user names), hashtags, unique ID, and permalink. To protect user privacy, all information related to user identity was discarded. Although many of these attributes may be useful for future models, the current study focuses only on the tweet content.

\paragraph{Preprocessing}
\label{sec:preprocessing}

To efficiently input tweet content into machine learning models in a way that is understandable, we first need to do preprocessing to clean the data and make it less noisy. This preprocessing step is a common practice in NLP models to ensure that the remaining word tokens are meaningful. Each tweet has gone through the following process in sequence: 

\begin{enumerate}

\item Converted all English alphabet characters to lower case. 
\item Removed all the URLs. 
\item Removed the symbols `\url{@}' and `\#'. 
\item Removed all the characters that are not in the English alphabet, to filter out numbers and non-English tweets using the library spaCy.
\item Removed sentences with only 1 word token left. 
\end{enumerate}

\begin{figure}
    \centering
\includegraphics[width=0.6\textwidth]{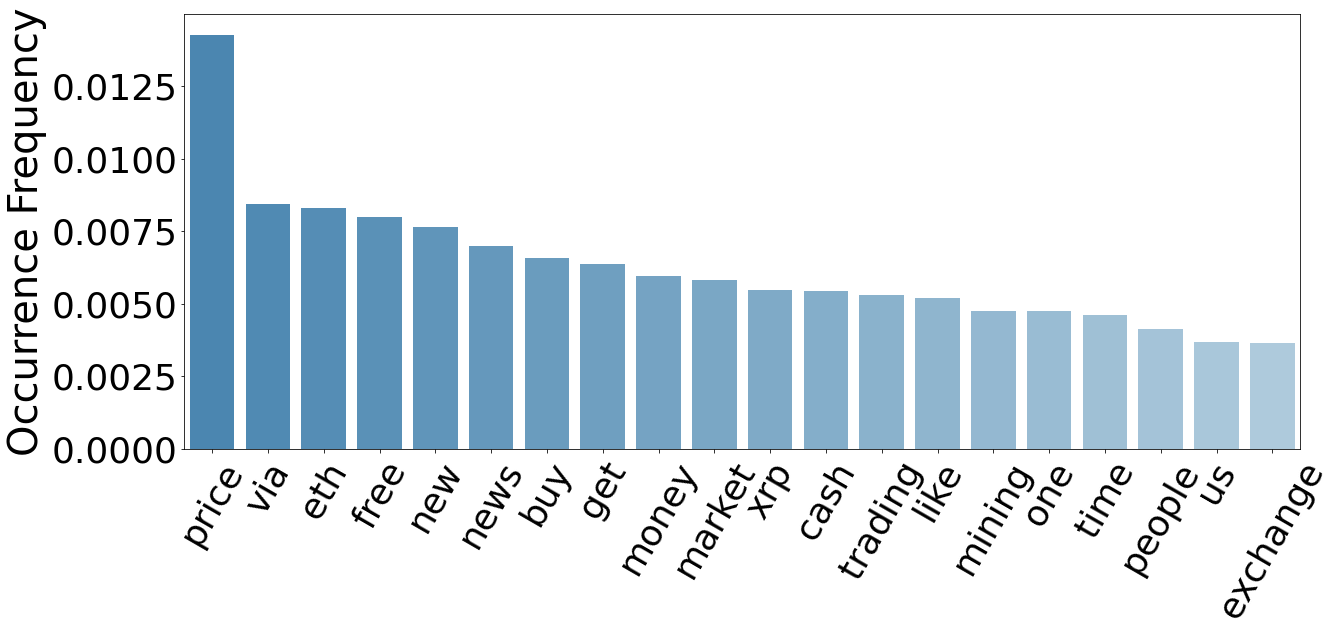}
    \caption{Top 20 most frequent words in the tweets in our dataset.}
    \label{fig:word_count}
\end{figure}

Figure~\ref{fig:word_count} illustrates the 20 words with the highest occurrence frequency from the entire Twitter dataset. There are a total 36,639 unique words. Stopwords, `bitcoin', `btc' and `cryptocurrency' have been excluded from the counting process, as unsurprisingly, they are the most frequent words given our search criteria when constructing the dataset. We notice two other cryptocurrencies were often mentioned together, `eth' (Ethereum) and `xrp'( Ripple). Hence, we proceeded to include Ethereum price as part of the technical indicators for TA dataset. Action words such as `buy' and `get' also occurred with a high frequency.

\subsection{TA Data}
For simplicity, we refer to the price related input data as TA data. It consists of three elements: candlestick data (Open-High-Low-Close-Volume, or OHLCV), related asset prices, and a few selected technical indicators. We will discuss each of these in more details below. 

\paragraph{Candlestick data} We included the daily Bitcoin OHLCV data from CryptoCompare\footnote{\url{cryptocompare.com}}. As Bitcoin is traded on multiple exchanges, data from one exchange may not capture the full picture. Cryptocompare aggregates the trading volume and prices from different exchanges to provide a more comprehensive overview of market activities (also used by \citet{alonso2020convolution}). The data covers the period from 1 January 2015 until 31 May 2021, which is the same range as the collected Twitter data.  


\paragraph{Technical Indicators and correlated assets}
\label{sec:ti}

\begin{figure*}
\centering
\begin{subfigure}{0.49\textwidth}
\centering
\includegraphics[width=1\textwidth]{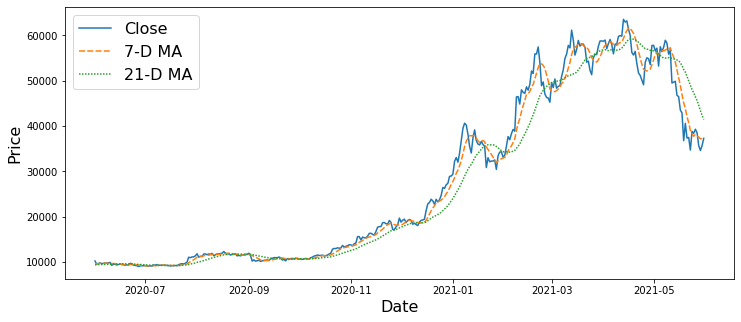}
\caption{BTC Close price with 7-D and 21-D MA.}
\label{fig:Close_with_ma}
\label{fig:my_label}
\end{subfigure}
\begin{subfigure}{0.49\textwidth}
    \centering
\includegraphics[width=1\textwidth]{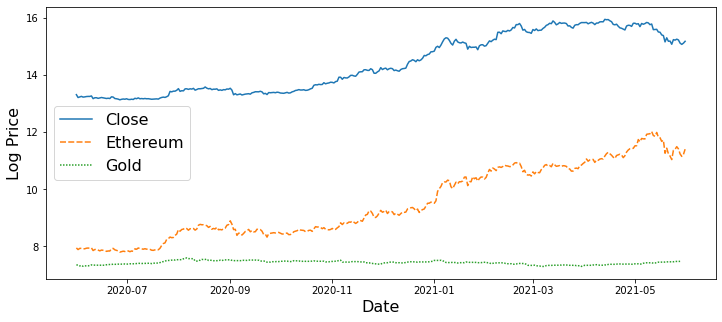}
    \caption{BTC Close price with Ethereum and Gold price on a log scale.}
    \label{fig:Close_with_gold_eth}
\end{subfigure}
\begin{subfigure}{0.49\textwidth}
    \centering
\includegraphics[width=1\textwidth]{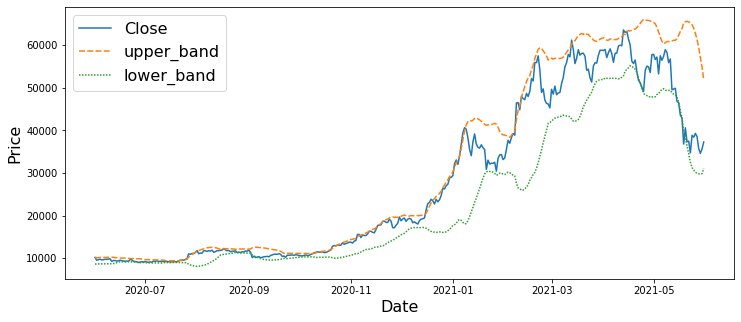}
    \caption{BTC Close price with Bollinger bands.}
    \label{fig:Close_with_bollinger}
\end{subfigure}
\begin{subfigure}{0.49\textwidth}
    \centering
\includegraphics[width=1\textwidth]{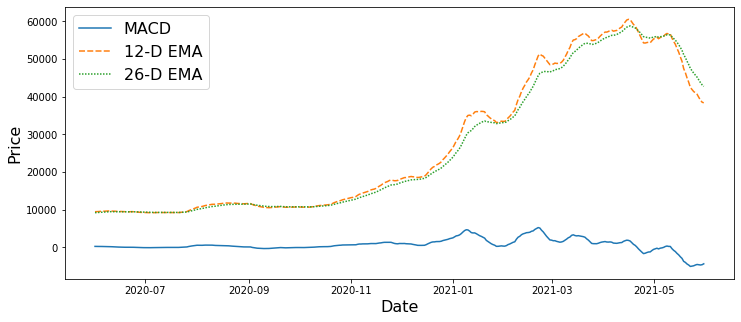}
    \caption{MACD with 12-D and 26-D EMA.}
    \label{fig:MACD_with_ema}
\end{subfigure}
\caption{Visualisation of selected TA features.}
\label{fig:price_plots}
\end{figure*}

In addition to the basic OHLCV data collected directly from Cryptocompare, we have also calculated 13 standard technical indicators, including correlated asset prices. Figure~\ref{fig:price_plots} visualizes these indicators together with the Bitcoin close price. For better visibility, only the last year of our data is displayed.


\begin{itemize}
    \item Moving Averages (5) - Moving average is a commonly used feature in technical analysis \citep{ellis2005smarter}. We have included five different moving averages: the 7-day simple moving average, the 21-day simple moving average, and three exponential moving averages. The first exponential moving average uses a decay rate of 0.67. To support the calculation of Moving Average Convergence Divergence (MACD), we calculated 12-day and 26-day exponential moving average and kept them as indicators.  
    \item Moving Average Convergence Divergence (MACD) (1) - this indicator is built upon moving averages. It compares the short-term moving average to the long-term moving average in order to identify the price movement momentum. If the short-term moving average is greater than the long-term moving average, it suggests that the recent price demonstrates an upward momentum. In our set-up, we have selected the 12-day and 26-day exponential moving averages to calculate the MACD. 
    \item 20-day Standard Deviation of BTC Closing price (1)- this is a basic measure of the BTC price volatility, and used to calculate the Bollinger Bands. 
    \item Bollinger Bands (2) - Bollinger Bands are volatility bands placed above and below the moving average of price. We have set up the band to be $\pm$ two 20-day standard deviation of the price from the 21-day simple moving average. The band captures information on price volatility.  
    \item High-Low Spread (1) - This is the distance between the highest  and lowest price of the day. The indicator attempts to capture the price volatility of the day. 
    \item ETH price (1) - the close price of Ethereum on the same day.  Bitcoin and Ethereum currently the two cryptocurrencies with the two largest market caps (excluding USDT). Their price has historically shown correlation \citep{katsiampa2019volatility, beneki2019investigating}. 
    \item Gold spot price (1) - Bitcoin is often referred to as a popular inflation hedge, or `digital gold' \citep{kang2019co}, hence we have included the Gold price. 
    \item Moving Average Indicator (1) - This feature is a binary representation which indicates whether the 7-day simple moving average price of the day is 5\% higher than the current price. 
    
\end{itemize}


\paragraph{Normalising Procedure}

The features that are directly related to the Bitcoin price were normalised as percentage change of the closing price of the previous day as per Equation~\ref{eq:normbtc}. These include OHLC, moving averages, and Bollinger Bands. Other features including volume, ETH price and gold spot price were normalised as percentage change over their own value of the previous day as per Equation~\ref{eq:normself}; Lastly, for MACD, 20-day standard deviation and high-low spread, we normalised as percentages of the closing price of the previous day, as per Equation~\ref{eq:normself2}. 

\begin{equation}
    \text{feat\_norm\_btc}_t = \frac{\text{feat}_{(t)}-\text{price\_BTC\_close}_{(t-1)}} {\text{price\_BTC\_close}_{(t-1)}}
    \label{eq:normbtc}
\end{equation}

\begin{equation}
    \text{feat\_norm\_self}_t = \frac{\text{feat}_{(t)} -  \text{feat}_{(t-1)}}{\text{feat}_{(t-1)}}
    \label{eq:normself}
\end{equation}

\begin{equation}
    \text{feat\_norm\_prev}_t = \frac{\text{feat}_{(t)}}{\text{price\_BTC\_close}_{(t-1)}}
    \label{eq:normself2}
\end{equation}

\begin{figure*}[b]
    \centering
\includegraphics[width=0.9\textwidth]{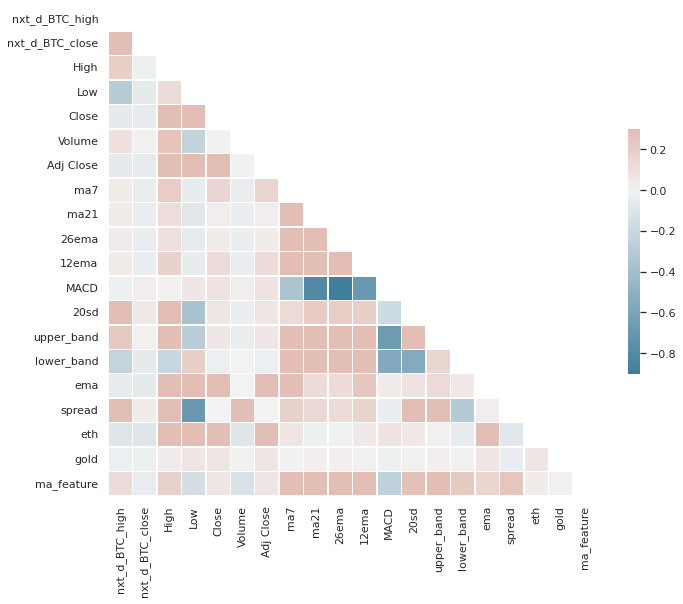}
    \caption{Correlation matrix for the TA indicators.}
    \label{fig:corr_matrix}
\end{figure*}

The Pearson correlation between the above mentioned (normalized) technical indicators and the next day Bitcoin price is shown in the Figure~\ref{fig:corr_matrix} and Table~\ref{tab:corr}. We notice that there is generally a low direct correlation between the features and the next day Bitcoin close price (normalized). The price of Ethereum has the highest correlation in terms of absolute value to the next day Bitcoin price. Volatility related indicators such as the 20-day standard deviation and lower Bollinger band show a stronger correlation as well. No one feature has an outspoken higher correlation with our predictive feature, hence we include all of them in our model. The full correlation values are shown in Table \ref{tab:corr_matrix_tab} in Appendix, and the corresponding p-values in Table \ref{tab:corr_matrix_pvalue} in Appendix. It is worth noting that, although the correlation values are low, the p-values for most of the features are also rather low. This shows that despite the low correlation in absolute terms, many features do still have statistically significant linear correlation to the next-day Bitcoin price. It is also worth mentioning that the features are later used in SVM models which are not linear models.

\begin{table}[h]
    \caption{Pearson Correlation coefficient of each feature (and p-value) with Bitcoin's next day close price.}
    \label{tab:corr}
    \centering
    \begin{tabular}{ccccc}
    \toprule
          \textbf{High} & \textbf{Low} & \textbf{Close} & \textbf{Volume}  \\
         \midrule
         -0.017(0.405) &-0.058(0.005) &-0.057(0.006) &0.013(0.497)\\
         \midrule
         \textbf{Adj Close} & \textbf{ma7} & \textbf{ma21} & \textbf{ema} \\
         \midrule 
         -0.056(0.06) &-0.039(0.054) &-0.032(0.098)&\emph{-0.060}(0.004) \\
        \midrule
        \textbf{26ema} & \textbf{12ema} & \textbf{MACD} & \textbf{20sd}  \\ 
        \midrule
        -0.034(0.084) &-0.037(0.060) &0.023(0.217) & 0.051(0.017)\\
        \midrule
        \textbf{upper band} & \textbf{lower band} & \textbf{spread} & \textbf{ma indicator} \\ 
        \midrule
        0.008(0.784) &\emph{-0.065}(0.002)  & 0.033(0.120)&-0.040(0.041)\\
        \midrule
        \textbf{eth} & \textbf{gold}  \\
        \midrule
        \emph{-0.092}(0.000) &-0.020(0.322)  \\
        \bottomrule
    \end{tabular}
\end{table}


\subsection{Data Labels}
Our research aims to answer whether we can predict extreme price movements using a hybrid multimodal model. Therefore, the prediction class labels should serve to flag out extreme price movements. 
We created binary class labels based on whether the percentage change of the daily BTC close price is above a selected threshold $\theta$ for each experimental cases. The digital asset market generally has much higher volatility over the traditional market, therefore, we chose the work with the daily `high' and `low' values respectively for the up and down task. We explored four values for the threshold $\theta$: $\pm2\% \text{ and } \pm5\%$. Being able to flag such large price movements may offer investors a warning sigh protect their assets from downside price movements or profit from upcoming large price movements. 

The $\pm5\%$ thresholds will serve to alert investors of upcoming extreme price movements. Since these are extreme events,  the positive and negative class ratio distribution ratio is around 1:5. With the $\pm2\%$ thresholds, the class distribution ratio is around 2:3, which is a lot more balanced. Although a 2\% move for Bitcoin might not be considered large given the market volatility, the more balanced dataset distribution case could serve to provide additional perspective to our experiments, which will be discussed later in Section~\ref{sec:experiments}. The distribution of the labels can be found in Table~\ref{tab:class_dist}. 

\begin{table*}[h!] \footnotesize
\caption{Class distribution for different predictive tasks $\theta$.}
\label{tab:class_dist}
    \centering 
    \begin{tabular}{l|ccc|ccc|ccc|ccc}
        \toprule
        \multirow{2}{*}{$\theta$} & \multicolumn{3}{c}{+5 \%} & \multicolumn{3}{c}{+2\%} & \multicolumn{3}{c}{-5 \%} & \multicolumn{3}{c}{-2\%} \\
        \cmidrule{2-4} \cmidrule{5-7} \cmidrule{8-10} \cmidrule{11-13} 
        {} & T & F & True Ratio & T & F &  True Ratio & T & F & True Ratio & T & F & True Ratio \\
        \midrule
        Training Set& 292  & 1680 & 14.84\%  & 810 & 1162 & 41.08\% & 298 & 1674 & 15.11\% & 789 & 1183 & 40.01\%  \\ 
        Test Set & 60  & 305 & 16.44\%  & 168 & 197 & 46.03\% & 60 & 305 & 16.44 \% & 175 & 190 & 47.95\%  \\
        \bottomrule

    \end{tabular}
\end{table*}

\section{Multimodal Model for Price Extreme Movement Prediction}
\label{sec:model}
\begin{figure*}[b]
    \centering
\includegraphics[width=0.9\textwidth]{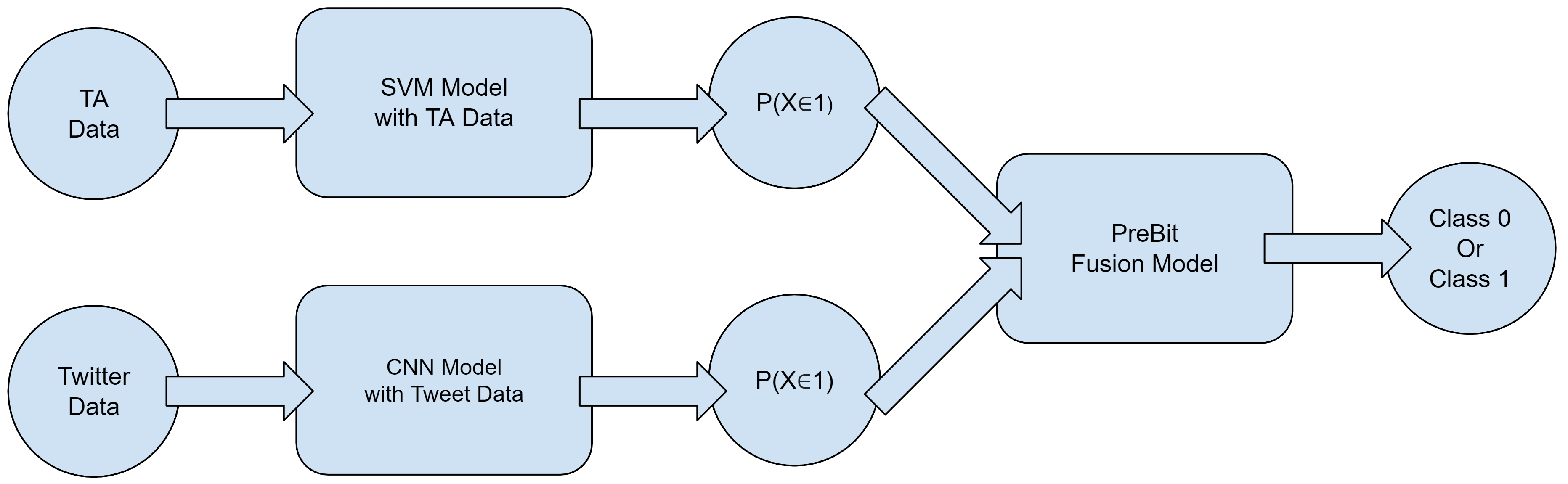}
    \caption{Proposed multimodal hybrid model. }
    \label{fig:architecture_3}
\end{figure*}

We propose a multimodal hybrid model that consists of two input modalities: Twitter content and TA data. A predictive model was built for each of these modalities and a fusion model was then added to the architecture to come to a joint prediction. The task of this model is to predict extreme price movement (up/down 2 or 5\%) within the next one day (i.e. looking at high/low prices, not just closing price). Figure~\ref{fig:architecture_3} shows an overview of the proposed architecture. We will discuss the three parts of our model in the next subsections. 

\subsection{CNN Model \myadded{with Tweet Data}}

For the \myadded{CNN} model\myadded{s with tweet data}, we first represented the daily tweets as a dense embedding, and then we implemented two possible convolutional neural network (CNN) configurations which were evaluated in the results section. \myadded{In the following text we will refer to the CNN models with tweet data as `Twitter CNN'.} 

\subsubsection{FinBERT context embeddings}

After the tweets are preprocessed (see Subsection~\ref{sec:preprocessing}), we represent them with efficient dense embeddings. There are various ways to obtain word embeddings, including Bag of Words (BoW), Term Frequency - Inverse Document Frequency (TF-IDF), Word2vec, Global Vectors for Word Representation (GloVe) \citep{pennington2014glove}. Earlier embedding models such as BoW and TF-IDF in essence create a sparse matrix based on the (co-occurrence) frequency of words. GloVe \citep{pennington2014glove}, a later model, uses a non-parametric and unsupervised learning algorithm that tries to capture the semantic similarities between words, and outputs them into a fixed-dimension feature representation. Word2vec embeddings are also much more dense and trained unsupervised by a simple neural network with one hidden layer \citep{mikolov2013distributed}.

We opted to use the context (sentence) embeddings from the state-of-the-art FinBERT model \citep{araci2019finbert} which is built upon Bidirectional Encoder Representations from Transformers (BERT) \citep{devlin2018bert} to obtain our embeddings. When BERT was developed in 2018 it outperformed traditional embedding models on variety of language tasks \citep{devlin2018bert}. FinBERT is in essence a  BERT model, finetuned on financial text datasets such as TRC2-financial\footnote{\url{https://trec.nist.gov/data/reuters/reuters.html}}, a dataset provided by Thomson Reuters, which consists of 1.8 million news articles published between 2008 and 2009. Because FinBERT is trained on financial texts, it is better suited for any financial related task such as the one tackled in this research. The FinBERT model obtained a 10 to 20\% increase in accuracy in Financial PhraseBank dataset classification tasks over the baseline models, and outperformed other state-of-the-art models in terms of both mean squared error (MSE) and $R^2$ on the FiQA sentiment dataset. These recent development incentivized us to use FinBERT in our model for embeddings.

The context embeddings for a given sentence generated by the FinBERT model are a 768-dimensional vector. As our dataset consists of 5,000 daily tweets across 6 years and 5 months, there would be over 9.5 million entries to process for the embeddings, resulting in a large amount (5,000) of embeddings per day. Considering the fact that tweets are very short (maximum 280 characters), and and with computational efficiency in mind, we decided to concatenate multiple tweets into one longer text. Specifically, one day worth of tweets is concatenate to form text slices of 200 word tokens, with 50 overlapping tokens at the start of next text slice. Finally the embeddings are obtained for each of the text slices and fed into the Twitter CNN model. Each embedding has a dimension of $1\times768$. Embeddings for tweets from the same day are stacked into a 2-dimensional vector of size $n \times 768$. Since the maximum number of text slices per day found in our dataset is 362, we used zero-padding on all the 2-dimensional input vectors to ensure a size of $362 \times 768$.  

\subsubsection{CNN Architectures}
\myadded{We opted to process the text with CNN architectures \citep{lecun1989handwritten}, as these have shown to be very efficient for two-dimensional data, as they have a greatly reduced number of parameters compared to traditional neural network structures.} We experimented with two different convolutional neural network (CNN) architectures for the language-based model: a sequential CNN and a parallel CNN.

\paragraph{Parallel CNN}

Applying CNNs for sentence classification has seen quite some research interest \citep{zhang2015sensitivity, zhang2016mgnc, hsu2017hybrid, shin2018contextual}. The popular work by \citet{kim-2014-convolutional} offered the basis for our model. In his work, Kim used the 300-dimensional Word2Vec embedding for each token in a sentence, and fed them into a 1-D CNN model. This work highlighted the importance of well-trained unsupervised pretraining of word vectors, and also demonstrated that using a simple 1-layer convolution can produce high performance on a variety of tasks.

In our work, we are focused on capturing the collective discussions and views on Bitcoin from the tweets on a given day. The information behind a singular tweet can often be trivial and noisy. Therefore, our model input consists of multiple embeddings which together capture a full day of tweet sentences as opposed to just a sentence of tokens in Kim's model. The intuition for this parallel CNN architecture is that the model should first capture the most relevant information between sentences from the embedding, and then extract the most relevant pieces of information from each of the 362 text slices (maximum number of daily text slices).

\begin{figure*}[h!]
    \centering
\includegraphics[width=0.95\textwidth]{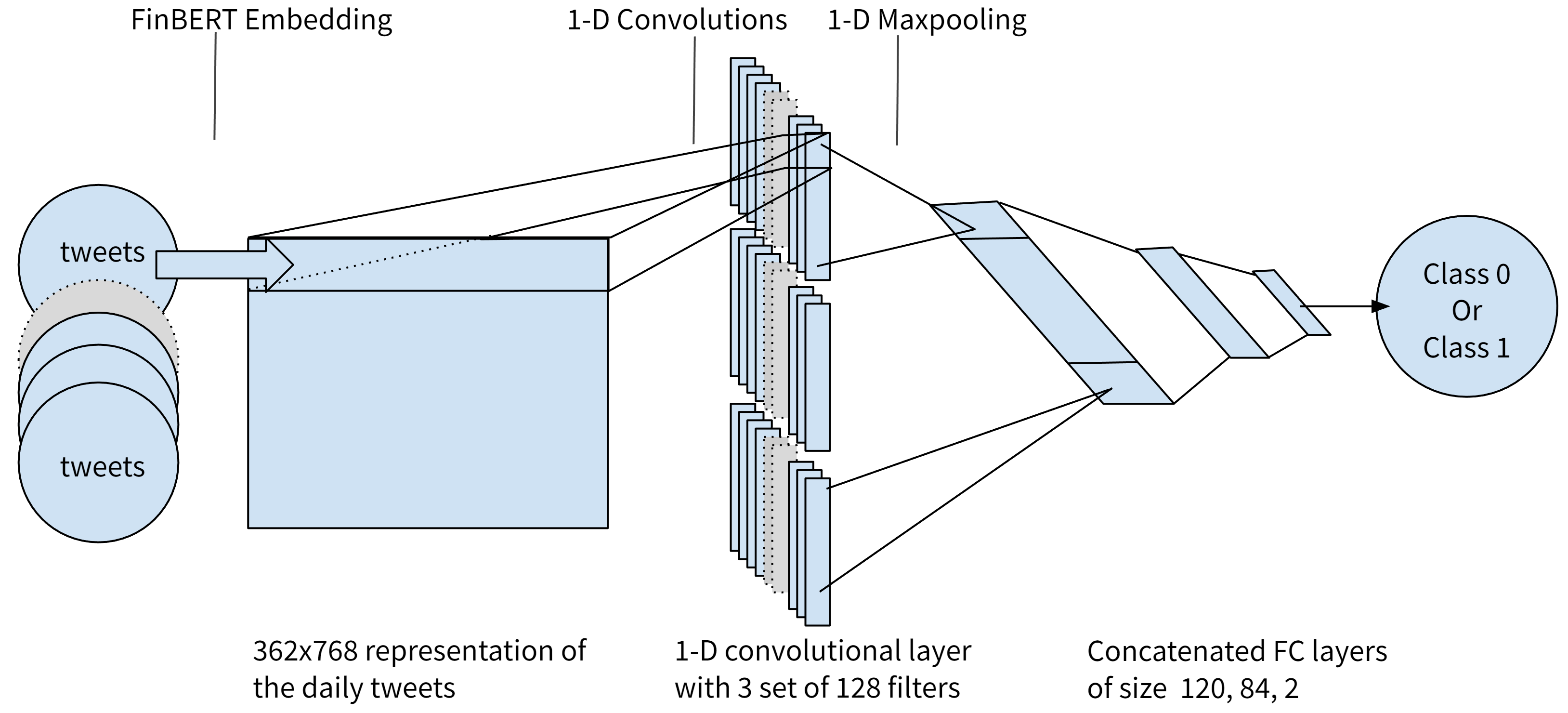}
    \caption{Parallel Twitter CNN model. We use FC for fully connected dense layer. A ReLu activation is applied after each convolutional layer. A dropout of 0.5 is applied on the first fully connected layer. The final prediction is made with softmax.}
    \label{fig:architecture_1}
\end{figure*}

Our proposed parallel CNN model first applies 1-D convolution on the input (embedding) layer. This convolution operation uses three sets of filters of size: 3$\times$768, 4$\times$768, and 5$\times$768, but the filters move only in one direction. Afterwards, we apply 1-D max pooling on the 3 sets of feature maps resulting from the convolution operations, with the feature map length as the kernel size (resulting in one value per feature map), and concatenate the output. Then we pass the result through two fully connected layers followed by the classification layer. An overview is provided in Figure~\ref{fig:architecture_1}.

\begin{figure*}
    \centering
\includegraphics[width=0.95\textwidth]{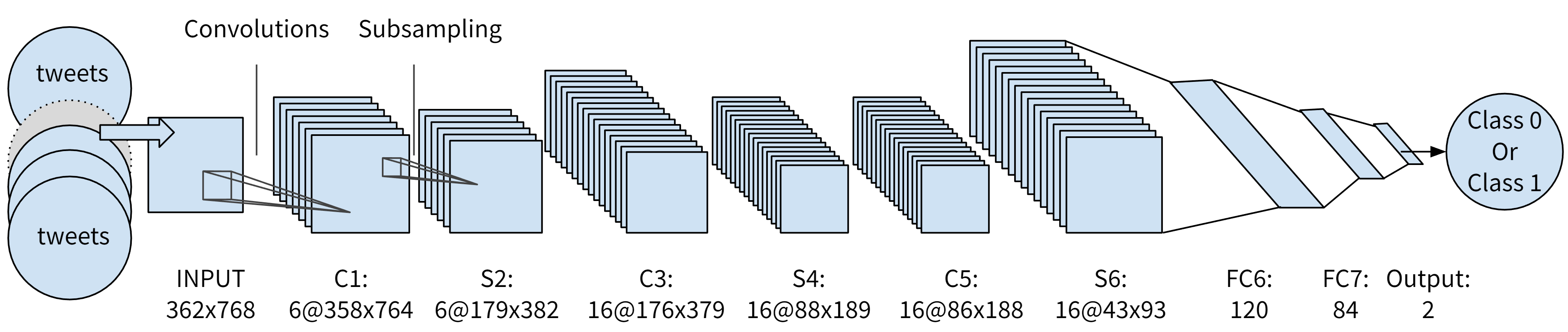}
    \caption{Sequential Twitter CNN model. We use Ci for convolutional layer i, Si for subsampling layer i, and FC for fully connected dense layer. A ReLu activation is applied after each convolutional layer, as well as after the last fully connected layer. A dropout of 0.5 is applied on the first fully connected layer. The final prediction is made with softmax. }
    \label{fig:architecture_2}
\end{figure*}

\paragraph{Sequential CNN}
A sequential CNN model (see Figure~\ref{fig:architecture_2}) was also implemented. This model essentially treats the embeddings as images of size $362 \times 768$ and applies a 2-D convolution operation on this input. This architecture is inspired by the LeNet-5 model~\citep{lecun1998gradient}. The main difference is that we added one extra convolutional layer and used a filter size of 5$\times$5, 4$\times$4, 3$\times$3 respectively in each of the three convolution layers. 

\subsubsection{Implementation and hyperparameters}

\paragraph{Loss function}
Two different loss functions were implemented for the Twitter CNN model: a conventional Cross Entropy (CE) loss and a Focal Loss \citep{lin2017focal} developed by Facebook AI Research. The CE loss is a staple in classification model training. Since our dataset is unbalanced (see Section~\ref{sec:dataset}, there is an incentive to use a loss function that takes this into account, i.e. Focal Loss (FL).


Focal Loss was proposed by \citet{lin2017focal} to train dense object detectors. Below we show the formulas for the two loss functions for each output sample, in the case of a binary classification task (class $t = 1$ or $t = 0$), with $p_t$ being the probability of a sample belonging to class $t$. 

\begin{equation}
       CE(p_t)=
           \begin{cases}
           -log(p_t)& \text{if } y=1\\
           -log(1 - p_t)& \text{otherwise} \\
           \end{cases}
\end{equation}
\begin{equation}
       FL(p_t)=-\alpha_t(1-p_t)^\gamma log(p_t)\\
      \text{  with  } p_t = 
           \begin{cases}
                p  & \text{if} t = 1\\
                1-p & \text{otherwise}
           \end{cases}
\end{equation}

As we can see, Focal Loss differs from cross entropy loss (CE) through the additional factor  $-\alpha_t(1-p_t)^\gamma$. The parameter $\alpha$ ranges from 0 to 1 and attempts to tackle the class imbalance directly by amplifying the loss from the minority class. It is usually set as the inverse class frequency or tuned on the validation set. 
The parameter $\gamma$ attempts to reduce the loss contributed by high confidence classifications, namely the easy examples, and generally is in the range from 0 to 5 to be effective. These parameters prevent the model from being overwhelmed by the easy negatives and enable the model to focus on the minority positives. \citet{lin2017focal} reported that detectors trained with FL showed superior accuracy results compared to state-of-the-art detectors trained with BCE loss. 

\paragraph{Hyperparameters}
The model was implemented in PyTorch, and we used the standard Adam optimiser with PyTorch default parameters. For the Sequential CNN model, we added L2 regularisation through the Adam optimizer's $weight\_decay$ parameter set at 0.0005 to 0.001 to prevent overfitting. A ReLU activation is applied after each convolutional layer. A dropout of 0.5 is applied on the first fully connected layer. Additionally, ReLU activations are also applied after fully connected layers in the Sequential CNN model.  

The performance of the sequential and parallel CNN models will be compared in the experiment section. We should note that the sequential CNN is computationally more expensive as the model has 7.6 million trainable parameters versus the 2.6 million in the parallel model, this may provide difficulties when training on small datasets. Yet, these layer, loss function, and kernel related hyperparameters were chosen based on best performance with trial-and-error on the validation set. The original training set was split is a (language) training set (90\%) and a (language) validation set (10\%). \myadded{The workflow used to develop (finetune, train, and test) the CNN models is illustrated in Algorithm \ref{alg:twitter_pseudo}.}

\subsection{ \myadded{SVM} Model \myadded{with TA data}}

\myadded{We implemented a Support Vector Machines (SVM) model \citep{cortes1995support} for making extreme price movement predictions using only the TA data. 
In the following text we will refer to the SVM model with TA data as TA SVM or TA model.} The TA model takes as input all available TA data as described in Section~\ref{sec:ti}, including candlestick data, and technical indicators including correlated assets, to predict extreme price movement (up/down 2-5\%). We experimented with several simple models \myadded{and compared their F1-scores (best F1-score in bracket for task up 5\%)}. These models include logistic regression \myadded{(0.79)}, support vector machines (SVM) (0.97), a 2-layered feed-forward neural network with dimensions 512 (0.90), as well as a 1-layered long-short term memory model (LSTM) with a hidden dimension of 256 \myadded{(0.86)}. From these test, we have selected the best performing model --SVM-- to use as part of our model. While this model may seem simple, its performance makes sense given the limited size of our dataset.

\paragraph{Model Input}
This model will take as input all of the features described in Section~\ref{sec:ti}: OHLCV data as well as 13 technical indicators, resulting in a total of 19 features per day. To provide additional historical price information, we concatenated this data in windows of 5 days. This resulted in a final input size of $1\times95$. 
We experimented with Principle Component Analysis (PCA) values to reduce the dimension of the input, however, it did not produce better results. Thus, in the final version of the model, no PCA was applied.

The \myadded{TA} SVM model was implemented with Scikit-learn. Based on trial-and-error, we opted to use the Radial Basis Function (RBF) kernel. The RBF kernel has 2 input parameters: C and gamma. We performed a Grid Search to determine the optimal C and gamma. This search used 4-fold cross validation with the F1-score as the evaluation metric to guide the search. 
The reason for using F1-score as the evaluation criteria will be discussed further in Section~\ref{sec:experiment}. \myadded{The workflow used to develop (finetune, train, and test) the SVM  model is shown in Algorithm \ref{alg:price_vol_pseudo}.}

\subsection{\myadded{PreBit} Fusion Model}

The proposed Fusion model takes the prediction probabilities output by the Twitter and TA model as input and makes a final prediction about the extreme price movement. An overview of the Fusion model can be found in Figure~\ref{fig:architecture_3}.

\paragraph{Model Input}
The input to the Fusion model consists of the probabilities for the positive class from the Twitter model and TA model. More specifically, for the TA SVM model, we applied a sigmoid function to the decision function output. For Twitter model, we took the model output after the softmax function. The resulting probabilities from both models were concatenated so they were of size $1 \times 2$. This small vector forms the input to the fusion model.

\paragraph{Model Architecture}
We experimented with several models such as feed-forward neural networks (FNN), logistic regression, SVM with RBF kernel and SVM with polynomial kernel. Except for the neural network models, who were implemented in PyTorch, we used Scikit-learn to implement all the models. Given the limited size of our input dataset, and the results from this trial-and-error experiment, we proceeded to use SVM in our final experiments. The parameter selection process was conducted similar to that of the TA SVM model. \myadded{The workflow used to develop (finetune, train, and test) the PreBit Fusion model is illustrated in Algorithm \ref{alg:fusion_pseudo}.}



\begin{algorithm}[h]
\caption{Workflow for finding the best Model (SVM) with TA data.}
\label{alg:price_vol_pseudo}
\begin{algorithmic}[1]
\For{Task $n \in \{+5, -5, +2, -2\}$}
\Comment{Ready data}
\State $ X, Y \gets \text{loadTADataset()} $
\State $ X_\text{train}, X_\text{test} \gets \text{ percentageSplit}(X, 85\%: 15\%) $
\State $ Y_\text{train}, Y_\text{test} \gets \text{ percentageSplit}(Y, 85\%: 15\%) $

\Comment Tune parameters of model
\State $ p_\text{SVM} = [c=[0.1, 0.5, 1, 10,30, 40, 50, 75, 100, 500, 1000],
\gamma =[0.01, 0.05, 0.07, 0.1, 0.5, 1, 5, 10, 50]]$
 \State modelList = \{SVM RBF kernel, 
 SVM polynomial kernel\}
 
    \For{model $m \in $ modelList}
             \State Grid Search with parameters $p_\text{SVM}$ using crossValidationSplit$(X_\text{train}, 4)$  
            \EndFor
            
             \Comment{Report best model}
       \State Select model $m_{\text{best\_ta}}$ with highest F1-score 
    \State Report metrics for prediction on $X_\text{test}$ made by $m_{\text{best\_ta}}$ trained on $X_\text{train}$ 

 \Return $m_{\text{best\_ta}}$
\EndFor
\end{algorithmic}
\end{algorithm}

\begin{algorithm}[h]
\caption{Workflow for finding the best CNN Model with Twitter data.}
\label{alg:twitter_pseudo}
\begin{algorithmic}[1]
\For{Task $n \in \{+5, -5, +2, -2\}$}
\Comment{Ready data}
\State $ X, Y \gets \text{loadTwitterDataset()} $
\State $ X \gets \text{textPreprocess}(X)$
\State $ X \gets \text{extractFinBERTembeddings}(X)$

\State $ X_\text{train}, X_\text{val}, X_\text{test} \gets  \text{ percentageSplit}(X, 76.5\%:8.5\%:15\%) $
\State $ Y_\text{train}, Y_\text{val}, Y_\text{test} \gets  \text{ percentageSplit}(Y, 76.5\%:8.5\%:15\%) $

\Comment Tune parameters of model
    \State modelList = \{Parallel CNN, Sequential CNN\}
    \State $ p_\text{loss} = [\alpha=[0.1 \text{ to } 1.0 ],\gamma =[0,1,2,3,4,5]]$
    \For{model $m \in $ modelList}
            \For{loss function $\in$ \{Cross Entropy Loss, Focal Loss$[p_\text{loss}]\}$}
               \State Train $m$ using $X_{\text{train}}$ and $Y_{\text{train}}$ 
                   
            \EndFor
             \EndFor

                 \Comment{Report best model}
            \State Select model $m_{\text{best\_twitter}}$ with highest F1-score on validation set ($X_\text{val}$ and $Y_\text{val}$) 
    \State Report metrics for prediction on $X_\text{test}$ made by $m_{\text{best\_twitter}}$ trained on [$X_\text{train}+X_\text{val}$, $Y_\text{train}+Y_\text{val}$] 
    
     \Return $m_{\text{best\_twitter}}$
\EndFor
\end{algorithmic}
\end{algorithm}


\begin{algorithm}[h]
\caption{Workflow for finding the best PreBit Fusion Model.}
\label{alg:fusion_pseudo}
\begin{algorithmic}[1]
\For{Task $n$ in $\{+5, -5, +2, -2\}$}
\Comment{Ready data}
\State $ x, Y \gets \text{loadTADataset()} $

    \State $P_\text{TA} \gets \text{getProbability}( m_{\text{best\_ta}})$
    \State $P_\text{Twitter} \gets \text{getProbability}( m_{\text{best\_twitter}})$

    \State $X \gets (P_\text{TA},P_\text{Twitter})$

\State $ X_\text{train}, X_\text{test} \gets \text{ percentageSplit}(X, 85\%: 15\%) $
\State $ Y_\text{train}, Y_\text{test} \gets \text{ percentageSplit}(Y, 85\%: 15\%) $

    \Comment Tune parameters of model
    \State modelList = \{Logistic regression, SVM RBF kernel, SVM polynomial kernel\}
    \For{model $m \in $ modelList}
            \If{$m = $ SVM} 
             \State Grid Search with parameters $p_\text{SVM}$ using crossValidationSplit$(X_\text{train}, 4)$  
            \ElsIf{$m = $ Logistic regression} 
            \State Train $m$ using $X_{\text{train}}$ and $Y_{\text{train}}$ 
            \EndIf
    \EndFor

    \Comment{Report best model}
    \State Select model $m_{\text{best\_fusion}}$ with highest F1-score 
    \State Report metrics for prediction on $X_\text{test}$ made by $m_{\text{best\_fusion}}$ trained on $X_\text{train}$ 
    \Return $m_{\text{best\_fusion}}$

\EndFor

    \Comment{Threshold experiment for Task +5\%}
    \If{$n=+5$} 
    \State thresholdList = \{$0.5,0.6,0.7,0.8,0.9,0.95,0.99$\}
    \State $T_\text{fusion} \gets \text{getDecisionFunction}( m_{\text{best\_fusion}})$
    \For{threshold $\tau \in$ thresholdList}
        \For{instance $\in X_\text{test}$}
        \If{f $T_\text{fusion} \ge \tau $}
        \State Label as positive prediction
        \Else
         \State Label as negative prediction
         \EndIf
         \State
        \Return Predictions and confusion matrix of $m_{\text{best\_fusion}}$ for $X_\text{test}$ 
        \EndFor
    \EndFor
    \EndIf

\end{algorithmic}
\end{algorithm}


\begin{algorithm}[h]
\caption{Workflow for backtesting the different models. }
\label{alg:backtest_pseudo}
\begin{algorithmic}[1]
\For{period in \{full period, bull period, bear period\}} 
\Comment Setup
    \State   strategyList= \{Buy and Hold, 7-D and 21-D MA Cross, TA SVM, Fusion model, Fusion model 0.95 threshold, Fusion model 0.99 threshold\}
    \State cash $\gets$ USD 100, BTC $\gets$ 0
    
\Comment Process buy signals
\For{strategy in strategyList}
\State signalList $\gets$ getBuySignals(strategy)
    \For{time $t$ in period}
        \State signal $\gets$ signalList($t$)
       \If{signal is buy and BTC position is nil}
            \State Buy BTC with 100\% of cash 
            \Comment Close positions after a day
       \ElsIf{BTC position opened at $t-1$}
            \State Sell 100\% of BTC  
       \Else
            \State No action
       \EndIf
    \EndFor
    \EndFor
    \State \Return Action history and strategy performance metrics
\EndFor
\end{algorithmic}
\end{algorithm}

\section{Experimental setup}
\label{sec:experiment}

We want to uncover which elements of our hybrid multimodal model contribute most to accurately predicting extreme BTC price movements. In particular, we are interested to explore if the model accuracy improves by incorporating a predictive model based on Twitter data in our hybrid architecture. To properly explore this question we have performed an ablation study for each of our four tasks: will the BTC price go up by 5\%, down by 5\%, up by 2\%, and down by 2\% on the next day. Separate models were trained for each task. 

For each task, five models were evaluated in an ablation study to determine which input modality has the potential to improve predictions, and which CNN architecture is most efficient. These five models that were compared are the:
\begin{itemize}
    \item TA SVM model;
    \item Twitter \myadded{CNN} model (parallel);
    \item Twitter \myadded{CNN} model (sequential);
    \item Fusion model (parallel) - using output from parallel CNN model as part of the input; and the
    \item Fusion model (sequential) - using output from sequential CNN model as part of the input.
\end{itemize}

We should note that we cannot compare our model directly to other existing models as they typically use proprietary data sources, and do not always have their source code available. In addition, the published results of existing models are based on different time frames, hence they are not directly comparable. 
\myadded{To address this issue, we have created two random baseline models. }\myadded{We ran simulations for 1,000 times and reported their average and the 95\% confidence interval of the performances with the above mentioned five models in Table~\ref{tab:up5},~\ref{tab:up2},~\ref{tabledown5}, and~\ref{tab:down2}}:
\begin{itemize}
    \item \myadded{A uniformly random model that predicts class 1 half of the time, and otherwise class 0;}
    \item A stratified model that predicts class 1 and 0 according to their class distribution in the test set.
\end{itemize}

\subsection{Training-Test Split}
The dataset includes data from 1 January 2015 until 31 May 2021, and consists of a total of 2,337 entries after clean-up. For our experiments, we split the data into a training set and a test set. The latter consists of the last 1 year of our data, i.e. from 1 June 2020 until 31 May 2021, totalling 365 entries. This makes the split ratio around 5.4:1. The Bitcoin price during the test period includes a long bull run as well as a bear period (see Figure~\ref{fig:test_period}). Thus, we believed this to be as a good period to evaluate our model on. The non-stationary nature of the data makes it hard to generalize results, hence, we selected a test period with both types of markets to compensate as best as possible \citep{de2018advances}. All of the hyperparameter tuning was done only on a validation set which was part of the original training set (see Table~\ref{tab:class_dist}). For the Twitter CNN model, the training set was further split into a training and validation set with 9:1 ratio. We used the validation set to select the optimal hyperparameters including: Focal Loss parameters - $\alpha$ and $\gamma$, number of filters, number of layers and dropout ratio. For the TA and Fusion SVM models, 4-fold cross validation (on the training set) was used during the Grid Search for parameter tuning.

\subsection{Evaluation metrics for predictive model}
A very commonly used metric for any classification task is prediction accuracy. Although this provides an overview of the model performance, for datasets with noticeable class imbalance such as ours, accuracy alone is not enough. Therefore, we also include precision, recall, as well as F1-score. The latter is a comprehensive measure that accounts for both precision and recall rate for the prediction output as it is their harmonic mean. This means that both false positives as well as false negatives are considered. In addition, we report the confusion matrices to provide more complete insight into misclassifications. 
In our case this is particularly important as traders may be more interested to be absolutely certain of their model's predictions, and care less about missed opportunities, depending on their risk appetite, and hence want to focus on maximizing precision. We will illustrate this further in the backtesting section, whose setup is explained in the next subsection. 
In addition, we also compare our model performance to that of the baseline models. To reduce the variance in the baseline model performance, we ran 1,000 simulations and show the average performance as well as the 95\% confidence interval for each of the aforementioned metrics in Tables~\ref{tab:up5},~\ref{tab:up2},~\ref{tabledown5}, and~\ref{tab:down2}.



\subsection{Trading strategy backtesting}
\label{sec:backtest_setup}


To further evaluate how our model signals may be helpful in reducing risk during trading, we propose simple long-only trading strategies and analyse them with backtesting. The backtesting period is the same as the test set coverage, from 1 June 2020 until 31 May 2021. We assume no leverage, no transaction fees, complete liquidity and instant transactions without slippage. 

Since our models performed best for Task 1 (predict up 5\% BTC price), we have focused our trading strategy on this model. Both the TA and Fusion model (sequential) showed higher performance than the other models, they were used to construct the trading signals. 

\paragraph{Trading strategy}
We implemented the following trading rules: if the model predicts a 5\% upward price movement for the next day, it flags a buy signal. We then buy 100\% of all the cash holdings at the closing price of the day. The holding period is always set to one day, i.e., we always sell at the closing price the next day after buying. We limit the strategy to perform only one action per day, either buy, hold or sell. When there are consecutive days of buying signals, we only buy and hold during the first day. The occurrence of the aforementioned situation is rare during our test period, thus it has limited impact on the performance.   

\paragraph{Baselines and metrics}
In addition to the TA SVM and Fusion model (sequential), we have included four other strategies for comparison in our backtesting: 
\begin{itemize}
    \item Buy and Hold - Buy on the first day and sell on the last. A commonly use baseline comparison. 
    \item 7-D and 21-D Moving Average (MA) Cross - Buy when the 7-D MA goes above the 21-D, sell when 7-D MA dives below the 21-D MA. Sometimes it is referred to as the `Golden Cross'. It is a classic trading strategy capitalising on momentum \citep{liu2021bitcoin}. 
    \item Fusion model (sequential) with 0.95 prediction threshold - This is a variation on our Fusion model (sequential). The original Fusion model predicts between two classes by comparing the probability for each class to the default threshold 0.5. If the model's output probability is greater than 0.5, the predicted class will be positive. In this variation, we have increased this threshold to 0.95, meaning that the model only predicts the positive class only if it has extremely high confidence. We explore the influence of this on reducing the risk of the trading strategy. 
    \item Fusion model (sequential) with 0.99 prediction threshold - Similar to the above model, but with threshold set extremely high at 0.99.  
\end{itemize}

To evaluate the backtesting results, we examine the following metrics: 
\begin{itemize}
    \item Profit \% - The percentage of profit made. This reflects the overall performance of the strategy during the period. 
    \item Sharpe Ratio - A risk-adjusted measure of the return \citep{sharpe1998sharpe}. The risk-free interest rate is assumed to be 0 in our calculation. 
    \item Sortino Ratio - A variation of the Sharpe ratio that only factors in the downside risk \citep{chaudhry2008efficacy}. 
    \item Max Drawdown \% - An indicator for downside risk over the full trading period. It measures the maximum observed loss of the portfolio.
    \item Win \% - The ratio of profitable trades. 
    \item Number of Trades - The number of trades made, which may be dependent upon the transaction costs. Note that in this low-frequency trading scenario, we omit the trading costs. For a more comprehensive analysis of more complex trading strategies this should be included in future work. 
\end{itemize}

In the next section, the various results of our experiments are discussed.

\section{Results}
\label{sec:experiments}

We ran a number of different experiments. First, an ablation study was conducted to examine the influence of the different parts of our proposed hybrid multimodal model on the prediction accuracy for each of the four tasks. Next, the best models were used to construct basic trading strategies for which we report the backtesting results and explore if they can be used to mitigate risk and exposure to volatility. When constructing the strategies, we also investigate the influence of probability thresholds on risk reduction.

\subsection{Influence of different modalities}

To explore the influence of each of the separate components of our multimodal model on the prediction accuracy, we performed an ablation study for each of the four predictive tasks. For each task, we report the best performing model in terms of the aforementioned evaluation criteria as well as the hyperparameters used in Tables~\ref{tab:up5},~\ref{tab:up2},~\ref{tabledown5},~\ref{tab:down2}. The tables also include the results for the random baseline models. 
The reported evaluation metrics include precision, recall, F1-score and overall accuracy on the test set. Given the class imbalance in our dataset, we pay special attention to the F1-score for positive class, weighted F1-score, the positive class precision as well as the recall rate. Positive class precision has practical significance because if a trading strategy is built upon the model predictions, higher precision indicates higher confidence which may be useful information when executing a buy or sell signal.  While a positive class recall rate means that the proposed trading strategy would capture as many buy signals as possible, thus resulting in a lower opportunity cost for missing out.

\begin{table*}[h!] \scriptsize
\label{tableup5}
\caption{Performance results for the Task Up 5\%. We use P for parallel, S for sequential. }
\label{tab:up5} \setlength{\tabcolsep}{2pt}
    \centering 
    \begin{tabular}{l|cc|cc|ccc|c}
        \toprule
        \multirow{2}{*}{Models} & \multicolumn{2}{c}{Precision} & \multicolumn{2}{c}{Recall} & \multicolumn{3}{c}{F1-score} & \multirow{2}{*}{Accuracy} \\
        \cmidrule{2-3} \cmidrule{4-5} \cmidrule{6-8} 
        {} & T & F & T & F & T & F & Weighted & {} \\
        \midrule
        TA SVM & \textbf{0.32}  & 0.85 & 0.22  & \textbf{0.91} & 0.26 & \textbf{0.88} & \textbf{0.78} & 71.23 \\ 
        Twitter CNN (P) & 0.20  & 0.86 & 0.47  & 0.62 & 0.28 & 0.72 & 0.65 & 59.22 \\
        Twitter CNN (S) & 0.18  & \textbf{0.93} & \textbf{0.95}  & 0.13 & 0.22 & 0.30 & 0.24 & 26.10\\
        Fusion model ({P}) & 0.31  & 0.89 & 0.48  & 0.78 & 0.37 & 0.83 & 0.76 & 73.42\\
        Fusion model (S)& 0.31  & 0.89 & 0.50  & 0.78 & \textbf{0.38} & 0.83 & 0.76 & \textbf{73.70} \\
        \midrule
        Random \myadded{baseline model} & 0.16   & 0.83  & 0.49   & 0.50  & 0.24 & 0.63  & 0.56  &49.91\\
        95\% confidence interval & 0.13-0.20  & 0.80-0.87 & 0.38-0.62  & 0.48-0.52 & 0.19-0.31 & 0.60-0.66 & 0.53-0.60 &46.30-53.97\\
        Stratified \myadded{baseline model}& 0.16  & 0.84 & 0.16  & 0.84 & 0.16 & 0.84 & 0.72 &72.49\\
        95\% confidence interval & 0.08-0.25  & 0.82-0.85 & 0.08-0.25  & 0.82-0.85 & 0.08-0.25 & 0.82-0.85 & 0.69-0.75 &69.86-75.34\\
        \bottomrule
    \end{tabular}
\end{table*}

\begin{table*}[h!]\scriptsize
\label{tableup2} 
\caption{Performance results for the Task Up 2\%.  We use P for parallel, S for sequential.}
\label{tab:up2}\setlength{\tabcolsep}{2pt}
    \centering
    \begin{tabular}{l|cc|cc|ccc|c}
        \toprule
        \multirow{2}{*}{Models} & \multicolumn{2}{c}{Precision} & \multicolumn{2}{c}{Recall} & \multicolumn{3}{c}{F1-score} & \multirow{2}{*}{Accuracy} \\
        \cmidrule{2-3} \cmidrule{4-5} \cmidrule{6-8}
        {} & T & F & T & F & T & F & Weighted\\
        \midrule
        TA SVM & 0.61  & 0.63 & 0.49  & \textbf{0.73} & 0.54 & \textbf{0.67} & 0.61 & 61.91 \\
        Twitter CNN (parallel) & 0.48  & 0.62 & 0.80  & 0.27 & 0.60 & 0.37 & 0.48 & 52.48 \\
        Twitter CNN (sequential) & 0.48  & 0.61 & \textbf{0.81}  & 0.25 & 0.60 & 0.36 & 0.47 & 51.08 \\
        Fusion model (parallel) & \textbf{0.62}  & \textbf{0.67} & 0.60  & 0.68 & \textbf{0.61} & \textbf{0.67} & \textbf{0.64} & \textbf{64.38} \\
        Fusion model (sequential)& 0.54  & 0.66 & 0.70  & 0.50 & \textbf{0.61} & 0.57 & 0.59 & 58.90  \\
        \midrule
        Random \myadded{baseline model} & 0.46   & 0.54  & 0.50   & 0.50  & 0.48 & 0.52  & 0.50  &50.01\\
        \myadded{95\% confidence interval} & 0.41-0.51  & 0.49-0.59 & 0.44-0.55  & 0.45-0.55 & 0.42-0.53 & 0.47-0.57 & 0.45-0.55 &44.66-55.07\\
        Stratified \myadded{baseline model}& 0.46  & 0.54 & 0.46  & 0.54 & 0.46 & 0.54 & 0.50 &50.30\\
        \myadded{95\% confidence interval} & 0.40-0.51  & 0.49-0.58 & 0.40-0.51  & 0.49-0.58 & 0.40-0.51 & 0.49-0.58 & 0.45-0.55 &45.20-55.07\\
        \bottomrule
    \end{tabular}
\end{table*}

\begin{table*}[h!]\scriptsize
\caption{Performance results for the Task Down 5\%.  We use P for parallel, S for sequential.}
\label{tabledown5}
\label{tab:down5}\setlength{\tabcolsep}{2pt}
    \centering
    \begin{tabular}{l|cc|cc|ccc|c  }
        \toprule
        \multirow{2}{*}{Models} & \multicolumn{2}{c}{Precision} & \multicolumn{2}{c}{Recall} & \multicolumn{3}{c}{F1-score} & \multirow{2}{*}{Accuracy}\\
        \cmidrule{2-3} \cmidrule{4-5} \cmidrule{6-8}
        {} & T & F & T & F & T & F & Weighted\\
        \midrule
        TA SVM & \textbf{0.40}  & 0.87 & 0.28  & \textbf{0.92} & \textbf{0.33} & \textbf{0.89} & \textbf{0.80} & 81.36\\
        Twitter CNN (parallel) & 0.20  & \textbf{0.94} &\textbf{ 0.90}  & 0.31 & \textbf{0.33} & 0.46 & 0.44 & 40.63\\
        Twitter CNN (sequential) & 0.14  & 0.81 & 0.38  & 0.53 & 0.20 & 0.64 & 0.57 & 50.27 \\
        Fusion model (parallel) & 0.37  & 0.87 & 0.28  & 0.90 & 0.32 & 0.88 & 0.79 & 80.27\\
        Fusion model (sequential)& \textbf{0.40}  & 0.87 & 0.28  & \textbf{0.92} & \textbf{0.33} & \textbf{0.89} & \textbf{0.80} & \textbf{81.37} \\
        \midrule
        Random \myadded{baseline model}  & 0.16   & 0.83  & 0.50   & 0.50  & 0.25 & 0.63  & 0.56  &49.99\\
        95\% confidence interval & 0.13-0.20  & 0.80-0.87 & 0.38-0.62  & 0.48-0.52 & 0.19-0.31 & 0.60-0.66 & 0.53-0.60 &46.30-53.97\\
        Stratified \myadded{baseline model} & 0.16  & 0.84 & 0.16  & 0.84 & 0.16 & 0.84 & 0.72 &72.49\\
        95\% confidence interval & 0.08-0.25  & 0.82-0.85 & 0.08-0.25  & 0.82-0.85 & 0.08-0.25 & 0.82-0.85 & 0.69-0.75 &69.86-75.34\\
        \bottomrule
    \end{tabular}
\end{table*}

\begin{table*}[h!]\scriptsize

\caption{Performance results for the Task Down 2\%.  We use P for parallel, S for sequential.}
\label{tab:down2}\setlength{\tabcolsep}{2pt}
\label{tabledown2}
    \centering
    \begin{tabular}{l|cc|cc|ccc|c}
        \toprule
        \multirow{2}{*}{Models} & \multicolumn{2}{c}{Precision} & \multicolumn{2}{c}{Recall} & \multicolumn{3}{c}{F1-score} & \multirow{2}{*}{Accuracy}\\
        \cmidrule{2-3} \cmidrule{4-5} \cmidrule{6-8}
        {} & T & F & T & F & T & F & Weighted\\
        \midrule
        TA SVM  & 0.55  & 0.56 & 0.43  & \textbf{0.67} & 0.49 & \textbf{0.63} & \textbf{0.56} & \textbf{56.71}\\
        Twitter CNN (parallel) & 0.50  & \textbf{0.67} & \textbf{0.91}  & 0.17 & \textbf{0.65} & 0.28 & 0.45 & 53.85\\
        Twitter CNN (sequential) & 0.44  & 0.42 & 0.66  & 0.23 & 0.53 & 0.29 & 0.41 & 43.34\\
        Fusion model (parallel) & \textbf{0.56}  & 0.57 & 0.46  & 0.66 & 0.50 & 0.62 & \textbf{0.56} & 56.44 \\
        Fusion model (sequential)& 0.54  & 0.56 & 0.48  & 0.62 & 0.51 & 0.59 & 0.55 & 55.34 \\
        \midrule
        Random \myadded{baseline model} & 0.48   & 0.52  & 0.50   & 0.50  & 0.49 & 0.51  & 0.50  &50.11\\
        95\% confidence interval & 0.43-0.53  & 0.48-0.57 & 0.45-0.55  & 0.46-0.55 & 0.44-0.54 & 0.47-0.56 & 0.46-0.55 &45.48-54.79\\
        Stratified \myadded{baseline model} & 0.48  & 0.52 & 0.48  & 0.52 & 0.48 & 0.52 & 0.50 &50.24\\
        95\% confidence interval & 0.43-0.53  & 0.47-0.57 & 0.43-0.53  & 0.47-0.57 & 0.43-0.53 & 0.47-0.57 & 0.45-0.55 &45.21-55.07\\
        \bottomrule
    \end{tabular}
\end{table*}




\myadded{Overall, the performance of the proposed models is higher than the random baselines.} We also we observe a better performance by the Fusion models, compared to the individual modality models, on the tasks related to upward price prediction compared to the TA SVM models, thus confirming the importance of incorporating Twitter content. For the Task Up 5\%, (Table~\ref{tab:up5}), the Fusion model (sequential) shows a higher positive class F1-score as well as a higher overall accuracy compared to the models based on individual modalities. 

\begin{figure*}[h!]
   \centering
\includegraphics[width=0.95\textwidth]{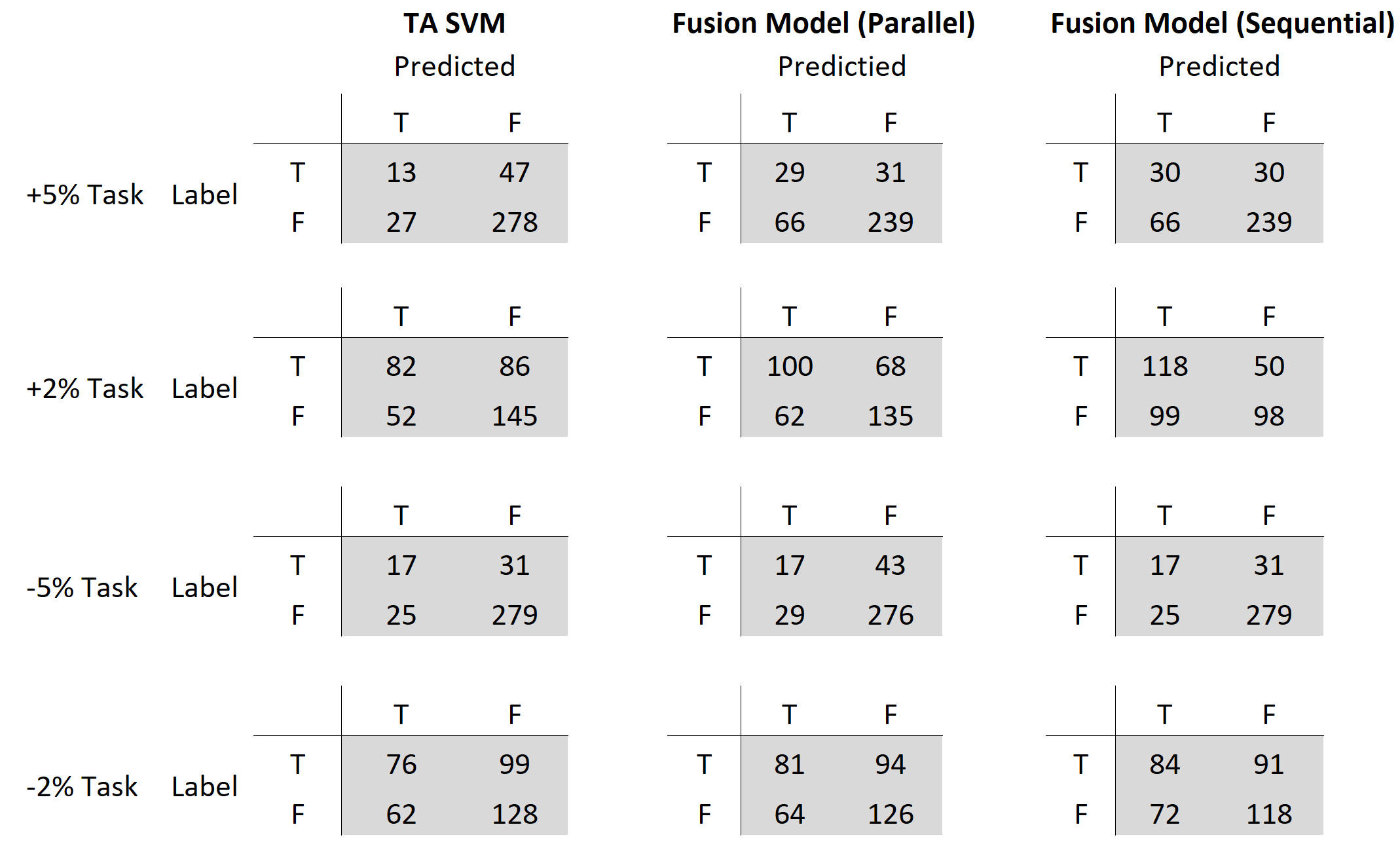}
    \caption{Confusion Matrices.}
    \label{fig:confusion_matrix}
\end{figure*}

\myadded{Looking at the precision/recall as well as the confusion matrices (Figure~\ref{fig:confusion_matrix}), we see that the SVM TA model is good at predicting true negatives, but misses a lot more true positives. In a trading scenario, we can interpret this as: the model may miss some opportunities as its predictions would be safer, more risk averse. This is the opposite of the models based on Twitter data, who perform the worst. It makes sense that their performance is less. In real-life, no trader would make trading decisions based purely on twitter information, without even glancing at the price data. The fusion models provide a balance between these two extremes, which is reflected in the higher F1 score as well as the confusion matrices in Figure~\ref{fig:confusion_matrix}.} For instance, the Fusion models were able to accurately predict around twice as many true positives for the Up 5\% Task, all the while maintaining the performance in terms of precision. 
From a practical point of view, this means that a trading strategy based on these signals may have twice as many winning trades and thus incur less opportunity cost due to staying market neutral. For Task Up 2\% (Table~\ref{tab:up2}), the good performance of the Fusion model (parallel) is even more apparent. \myadded{The improvements may be due to the fact that Fusion models are able to incorporate and capture more information than the individual models.} Except for negative class recall rate, all other metrics show improvements compared to the other models. Looking at Task Down 5 and 2 \% respectively (Tables~\ref{tab:down5} and~\ref{tab:down2}), the models have a comparable performance and the improvements due to the Twitter model are less obvious. This may be due to the fact that the TA SVM model is already quite good, arguably because of the strong correlation between Bitcoin price and some of the model inputs like Ethereum price and 20-day standard deviation of price (see Figure~\ref{fig:corr_matrix}). In future research, this effect may be increased by focusing on tweets by influencers in the `Crypto-Twitter sphere' instead of random tweets that mention Bitcoin, or by finetuning our word embedding representation to capture crypto- and Twitter-specific vocabulary.

While comparing the Fusion model performance to the average performance of the random baseline models, we observed better results across all tasks and in almost every evaluation metric. Since the stratified prediction model clearly outperforms the uniformly random prediction model, we focus on comparing our models to the stratified prediction model in the following discussion. The superior performance of Fusion models is clear for Task Up 2\% and Down 5\%. In these two tasks, the Fusion model outperforms the 95\% confidence interval (CI) upper bound of the stratified prediction model simulations in every metric. The performance is closer for Task Up 5\% and Down 2\%. In Task Up 5\%, even though the upper bound of the 95\% CI exceeds the Fusion model in terms of overall accuracy, it loses out by a large margin in precision, recall, and F1-score for the True class. And as mentioned in previous sections, these metrics are especially important to us due to the imbalanced class distribution of the dataset. Similarly in Task Down 2\%, the Fusion model's performance comes very close to the 95\% confidence interval upper bound of the stratified model, if not slightly better. Admittedly, it is not an easy task to prove that the results are \textit{statistically} significant. We have explored statistical tests such as the Diebold-and-Mariano test \citet{diebold2002comparing}. Unfortunately, the resulting p-values do not meet the criteria to reject the null hypothesis. We should, however, note that this test was designed to interpret regression forecasts, thus a lot less applicable to 0-1 classification classes. And as stated by \citet{diebold2015comparing}, it is not intended for model selection. Hence, we offer these results as is with the 95\% confidence interval, and in future research, we recognise that a more robust test should be performed on a larger test set.

When evaluating these models we should keep in mind the high class imbalance present in our dataset. In addition, since these predictions are quite directly translatable for use in a trading strategy, a resulting trading strategy could easily reach a win rate in the range of 50-60\%. Such rates are considered quite promising and may obtain good returns. A full analysis of a naive trading strategy is provided in Section~\ref{sec:backtest}. Overall, our proposed models have a good performance, with the upward extreme movement prediction models being successfully improved by adding Twitter models.



\subsection{Predictive threshold tweaking}

Before diving into an actual trading strategy that may be built upon our proposed models, it is important to consider the predictive threshold, especially since it has a direct influence on risk reduction. Our models were optimized based on F1-score, the harmonic mean of precision and recall rates, two metrics that can be traded off for one another by changing the predictive threshold.

The Fusion model's classification decisions are derived by comparing the SVM's output decision function to a threshold (0.5 as default). The parameters and decision functions for the SVM models have been optimized on the training dataset and fixed before getting the results on the testing set. Therefore, moving the threshold does not alter any model state, but only changes the reported classification results. If a higher precision rate (high-confidence positive class predictions) is desired, a higher threshold could be explored \citep{yu2015support}, as it only allows predictions with higher certainty to be classified as positive. 

\begin{figure*}[h!]
    \centering
\includegraphics[width=0.95\textwidth]{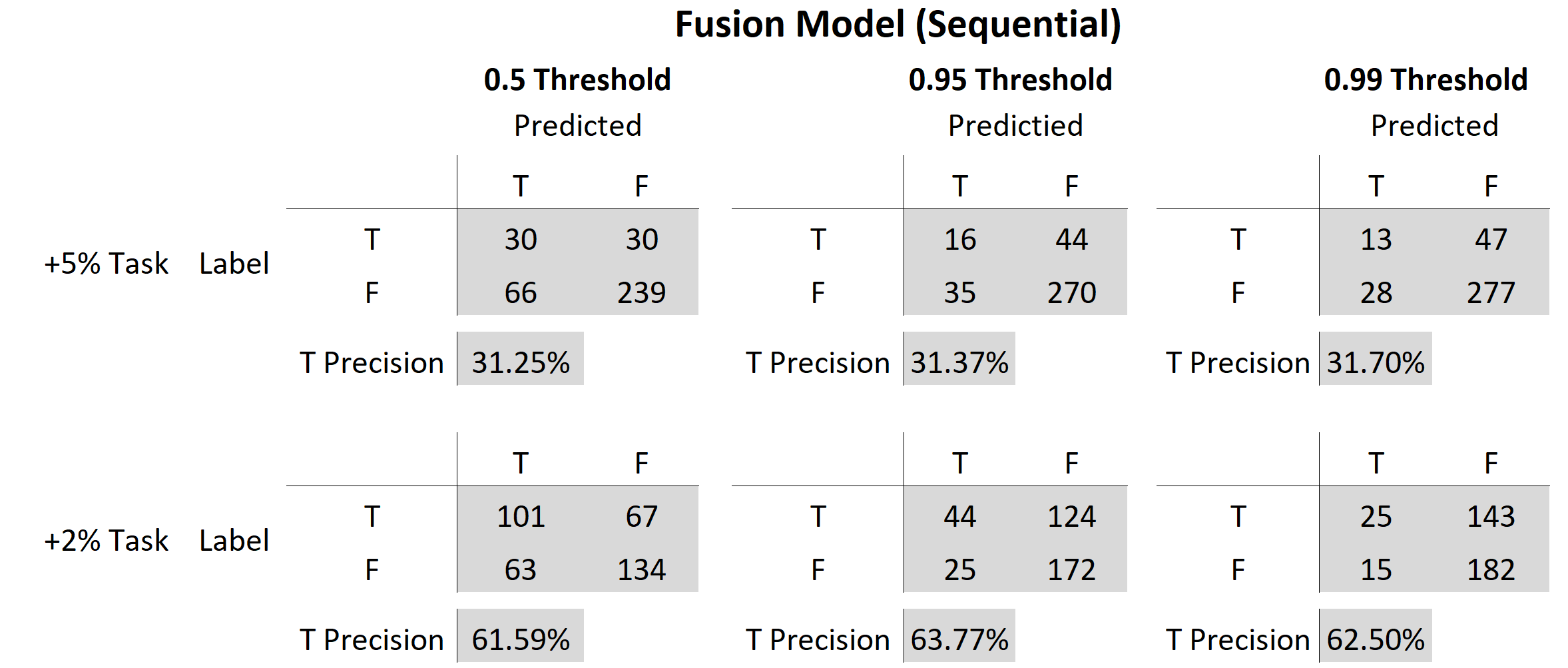}
    \caption{Confusion Matrices for Threshold Tuning.}
    \label{fig:confusion_threshold}
\end{figure*}

We report different confusion matrices based on a much higher decision threshold (0.95 and 0.99) in the best performing Fusion models for both Task Up 5\% and 2\% in Figure~\ref{fig:confusion_threshold}. For Task Up 5\%, the Fusion model (sequential)'s precision rate for the positive class improved slightly. The confusion matrices report a quite different amount of true positives. The improvement is more evident for the Fusion model (parallel) for Task Up 2\%, with a peak at the 0.95 threshold. 

In the next section, we will uncover the full impact of this threshold tweaking on the trading strategy. 




\subsection{Backtesting results}



\paragraph{Backtesting}
\label{sec:backtest}

\begin{table*}[htbp]

\caption{Backtesting results the full test period.}
\label{tab:backtest_overall}
    \centering
    \begin{tabular}{l|c|c|c|c|c|c}
        \toprule
        \multirow{2}{*}{Strategies} & \multirow{2}{*}{Profit \%} & \multirow{2}{*}{Sortino} & \multirow{2}{*}{Sharpe}  & \multirow{2}{*}{Max Drawdown \%}& \multirow{2}{*}{Win\%} & \multirow{2}{*}{Num of Trades}  \\
\\        \midrule
        Buy and Hold  & \textbf{249.3}  & 3.28 & 2.08  & 45.5 & N.A & N.A \\
        7-D and 21-D MA Cross & 199.7  & 3.60 & \textbf{2.17}  & 37.7 & 40.0 & 10 \\
        TA SVM & 60.4  & \textbf{3.62} & 1.56  & 16.0 & \textbf{58.0} & 31 \\
        Fusion model  & 1.4  & 0.26 & 0.18  & \textbf{12.4} & 56.0 & 50 \\
        Fusion model  0.95 threshold & 49.9  & 3.02 &  1.39  & 14.9  & 55.6 & 36\\
        Fusion model  0.99 threshold & 56.6  & 3.45 &  1.49  & 16.0  & 56.6 & 30\\
        \bottomrule
    \end{tabular}
\end{table*}

\begin{table*}[htbp]

\caption{Backtesting results the bull period.}
\label{tab:backtest_up}
    \centering
    \begin{tabular}{l|c|c|c|c|c|c}
        \toprule
        \multirow{2}{*}{Strategies} & \multirow{2}{*}{Profit \%} & \multirow{2}{*}{Sortino} & \multirow{2}{*}{Sharpe} & \multirow{2}{*}{Max Drawdown \%}& \multirow{2}{*}{Win\%} & \multirow{2}{*}{Num of Trades}  \\
\\        \midrule
        Buy and Hold  & \textbf{345.5}  & \textbf{8.10}  & \textbf{4.50} & 25.1 & N.A & N.A \\
        7-D and 21-D MA Cross & 96.0  & 4.16 & 2.43 & 29.9 & \textbf{75.0} & 5 \\
        TA SVM & 36.6  & 4.65 & 1.90  & 9.8 & 63.2 & 19 \\
        Fusion model & 12.9  & 1.72 & 1.09   & 12.4 & 57.1 & 28 \\
        Fusion model  0.95 threshold & 50.4  & 5.97 & 2.40 & \textbf{8.3}  & 68.2 & 22\\
        Fusion model  0.99 threshold & 33.4  & 4.32 & 1.77  & 9.9  & 61.1 & 18\\
        \bottomrule
    \end{tabular}
\end{table*}

\begin{table*}[htbp]

\caption{Backtesting results the bear period.}
\label{tab:backtest_down}
    \centering
    \begin{tabular}{l|c|c|c|c|c|c}
        \toprule
        \multirow{2}{*}{Strategies} & \multirow{2}{*}{Profit \%} & \multirow{2}{*}{Sortino} & \multirow{2}{*}{Sharpe} & \multirow{2}{*}{Max Drawdown \%}& \multirow{2}{*}{Win\%} & \multirow{2}{*}{Num of Trades}  \\
\\        \midrule
        Buy and Hold  & -40.5  & -4.27  & -3.33  & 45.5 & N.A & N.A \\
        7-D and 21-D MA Cross & -6.3  & -1.03 & -0.86  & 16.0 & 0.0 & 1 \\
        TA SVM & \textbf{32.0}  & \textbf{12.72} & \textbf{4.30} & \textbf{6.3} & \textbf{85.7} & 7 \\
        Fusion model  & -5.5  & -1.54 & -1.12  & 10.6 & 50.0 & 8 \\
        Fusion model 0.95 threshold & 17.0  & 6.96 & 2.89   & \textbf{6.3}  & 71.4 & 7\\
        Fusion model 0.99 threshold & \textbf{32.0}  & \textbf{12.73} & \textbf{4.30} & \textbf{6.3}  & \textbf{85.7} & 7\\
        \bottomrule
    \end{tabular}
\end{table*}



We implemented a simple, naive trading strategy based on the predicted classes for by our Fusion model as explained in Subsection \ref{sec:backtest_setup}. Since the best overall predictive results (in terms for F1-score) were obtained for predicting a 5\% increase in BTC price over the next day (see Table~\ref{tab:up5}), we have opted to further evaluate our models' performance for this task with backtesting.

The backtesting statistics were calculated using the library vectorbt \footnote{\url{https://github.com/polakowo/vectorbt}}. We report the results for our entire test period in Table~\ref{tab:backtest_overall}. Whenever we are evaluating financial data, the non-stationarity of the data poses limitations and skews the metrics \citep{de2018advances}. A rolling window test set may prove to be a slightly better representation, however, it would limit the amount of data we have to train our models. Hence, we ensured that our test period is long enough (1 entire year), and contains a steed upward (bull) as well as downward (bear) period. We report the results for the bull and bear periods separately in Tables~\ref{tab:backtest_up}, and~\ref{tab:backtest_down}.

Looking at the entire test period (Table~\ref{tab:backtest_overall}), a Buy and Hold strategy achieves the highest Profit \%. This is unsurprising given the rising trend in Bitcoin in the long run. This performance has a huge risk exposure, however, with a maximum drawdown (MDD) of 45.5\%. Not many investors would be willing to risk almost half of their capital before seeing gains. Hence, we explore how our models can be used to provide a more market neutral strategy with lower risk exposure. 

The Sortino ratio gives us an impression of risk free returns (excluding the upwards volatility, which is still included in the Sharpe ratio). Our proposed TA model achieves the highest Sortino ratio, followed closely by 7-D and 21-D MA Cross as well as the Fusion model with 0.99 threshold. The TA SVM model has a significanlty lower maximum drawdown of 16\% compared to the Buy and Hold drawdown of 45.5\%, while still obtaining a nice 60\% in returns with only 31 days of market exposure over the entire year. A similar result is obtained by the Fusion model with 0.5 and 0.99 threshold. 

When comparing the result of the three Fusion models, we observe consistently better results with the 0.95 threshold over the default threshold (0.5). Raising the threshold from 0.95 to 0.99, however, does not always yield enhanced performance. Threshold tweaking based on the strategy in use is essential. In future research, the impact of threshold selection based on custom, more advanced trading strategies could be investigated. We may also explore using different fusion techniques, such as early fusion. Overall, it would be good to test the generalisability and robustness of this model using more data, different market conditions, and assets. 

\begin{figure}
    \centering
\includegraphics[width=0.75\textwidth]{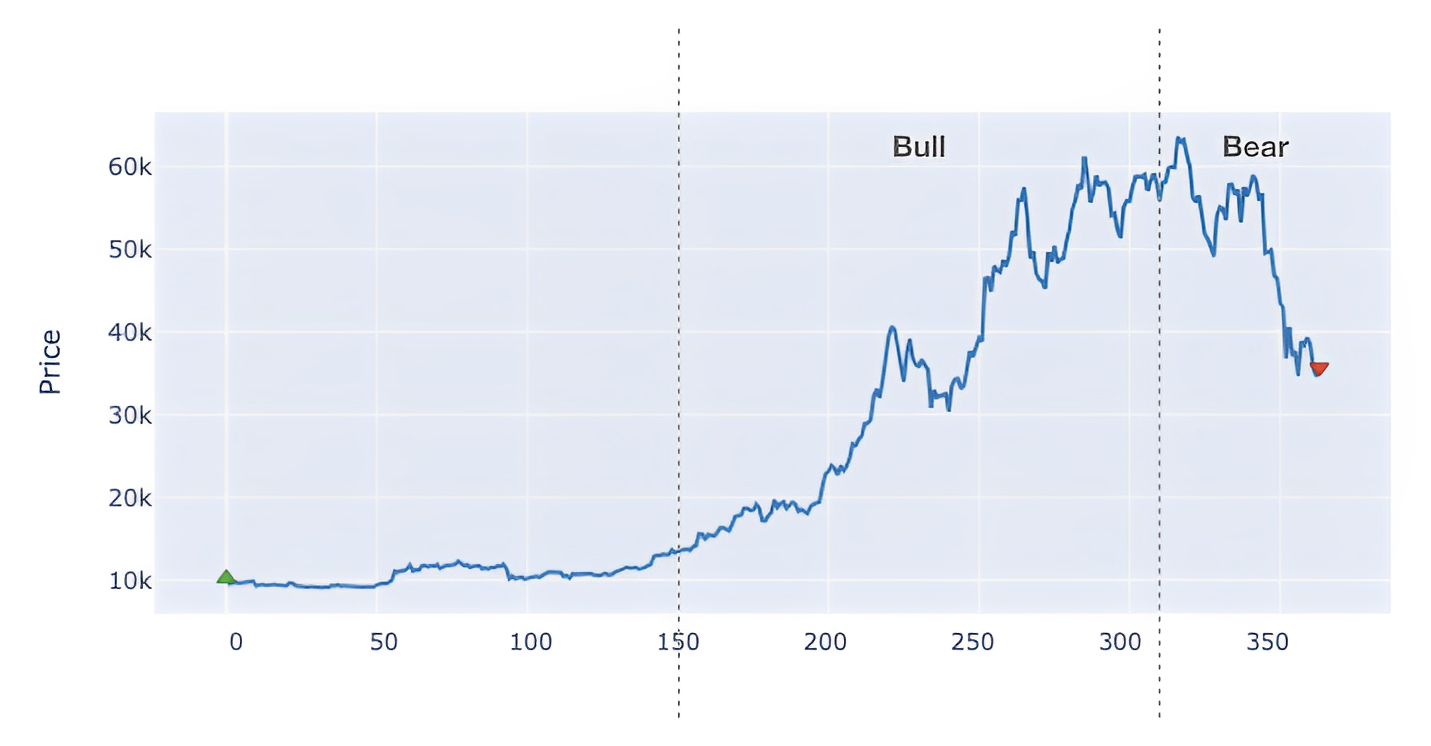}
    \caption{Bitcoin price during the entire test period.}
    \label{fig:test_period}
\end{figure}

Looking closer at the bull period (days 150-350) in Figure~\ref{fig:test_period}, we notice that, as expected, the Buy and Hold strategy performs very well, although still with a 25\% MDD. The proposed TA SVM and Fusion model with 0.95 threshold still achieve impressive performance in terms of Sortino ratio (4.65 and 5.97 respectively) and MDD (9.8\% and 8.3\% respectively), while sacrificing some profits for this reduced risk. 

It is during bear periods that the usefulness of our proposed models really becomes apparent. Table~\ref{tab:backtest_down} shows the trading results for the bear period (day 315 to day 365). During the bear period, both the TA SVM as well as the Fusion model with 0.95 and 0.99 threshold perform significantly better than the buy and Hold and the MA Cross benchmark models. While the latter is down as much as -40.5\% profit, the TA SVM and Fusion model with 0.99 threshold reaches 32\% profit. 
This nicely illustrates the usefulness of our proposed strategy, because while Bitcoin has seen tremendous growth, it also goes through extensive bear periods where risk management is essential. It is worth noting that this backtest is performed on a limited period in time, with a small number of trades, hence we could not perform a statistical significance test of these results. In future research, the model and trading strategy could be tested on different market conditions to further examine its robustness.


\begin{figure}
    \centering
\includegraphics[width=0.75\textwidth]{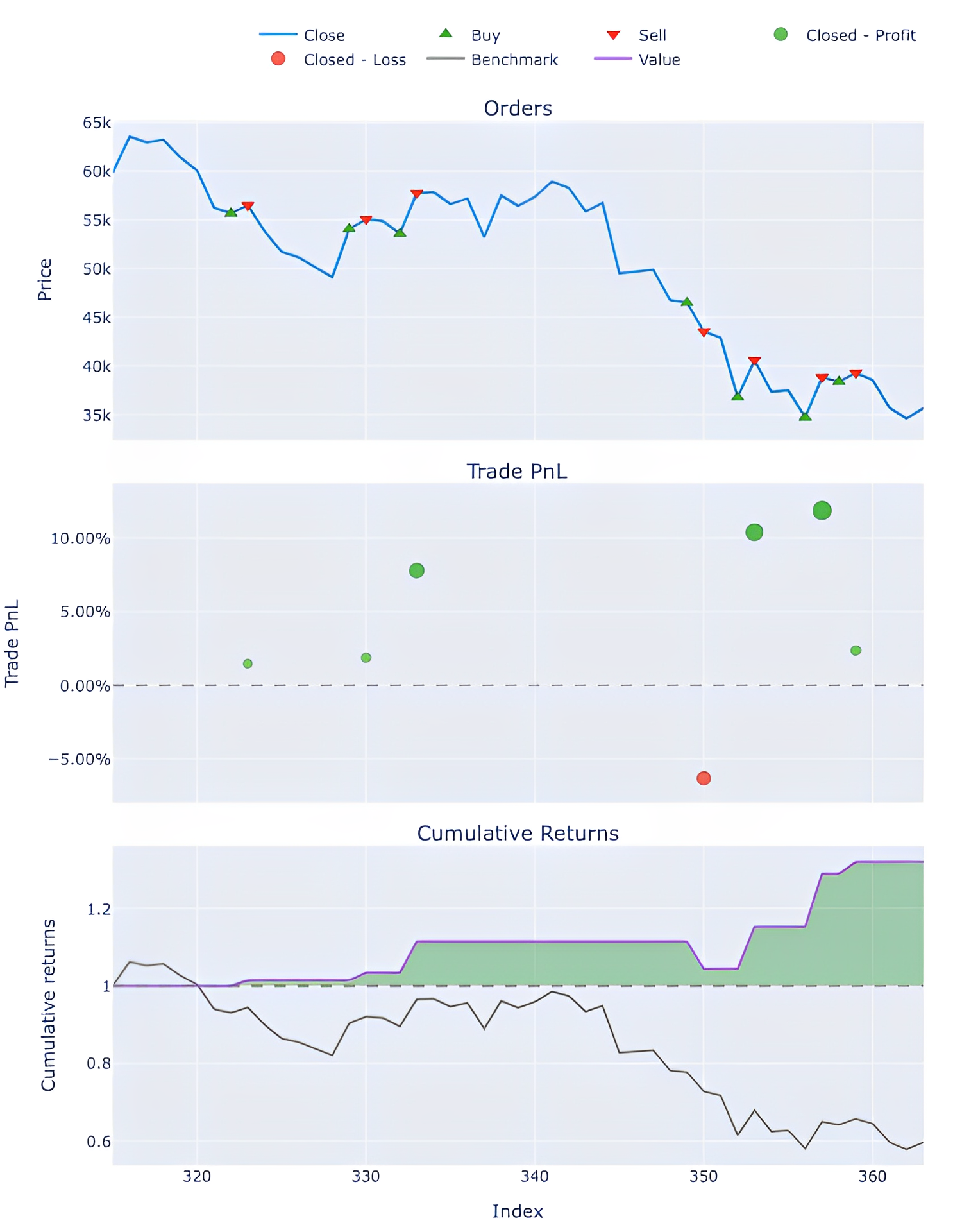}
    \caption{Trading strategy based on the TA model during the bear period.}
    \label{fig:bear_OHLCV}
\end{figure}

\section{Conclusion}

Bitcoin, and cryptocurrencies, are known for their volatile nature. We propose a cutting-edge multimodal model, PreBit, to predict extreme Bitcoin price movements (up/down 2 or 5 percent). In order to train our model, we created a new publicly available dataset, which includes 9,435,437 tweets that include the keyword `Bitcoin' from 1 Jan 2015 until 31 May 2021. We also included based candlestick price/volume data, as well as selected technical indicators and correlated asset prices (Ethereum and gold). The resulting multimodal ensemble model uses normalized data, as well as the finBERT context embeddings to provide a meaningful representation of our Twitter data. \myadded{The trained model and source code used in this manuscript is available online}\footnote{\url{https://github.com/AMAAI-Lab/PreBit}}.

In a thorough experiment, we perform an ablation study to compare the influence of adding the Twitter model or TA model to our hybrid model. This shows that adding prediction based on Twitter content improves the overall performance of the model for upward Bitcoin price prediction. Our proposed Fusion models demonstrate superior performance in positive class F1-score as well as overall accuracy in upward price prediction tasks compared to the TA SVM which uses only price and technical analysis data.

To further evaluate our model's performance and demonstrate its practical use, we propose a simple (long only) trading strategy and reported the backtesting results for our models that predict Up 5\% price movement. During this backtesting, we explored the influence of tweaking the predictive threshold on risk management. The results confirm the superior performance of our proposed TA SVM model as well as our multimodal Fusion model with 0.95 threshold in risk-adjusted measures such as Sortino ratio and maximum drawdown. While Buy and Hold strategies typically work well for Bitcoin and obtain huge profits, the risks can be substantial, with max. drawdown reaching 45.5\% in our test period. Our models substantially reduce this risk while maintaining an impressive profit ratio. 

The usefulness of our proposed approach becomes especially apparent during the bear market, when our Fusion strategy manages achieved 32\% Profit (with long positions only), despite the fact that the Bitcoin price was down by -40.5\%. We further observe that a carefully selected probability threshold can significantly improve the trading performance and lower the market exposure risk. 

However, as aforementioned, the evaluation and backtest is performed on a limited period in time. Potential threats to the model validity could be if the amount of tweets selected is not representable of the day. With more and more posts being made about Bitcoin, 5000 may not be a large enough sample to capture the entire market sentiment. Selecting these tweets from famous Bitcoin influencers may provide a remedy for this as they are seen and liked by a large number of followers. \myadded{Recent work by \citet{otabek2022twitter} confirms that tweets by users with the a high leverl of followers consequently have a influence on a future BTC price.} In addition, the Bitcoin market is very recent. As more price history builds up, predictions will get more and more accurate. Given the non-stationarity of such price series, we should also consider that we could have coincidentally taken a good or bad period concerning model accuracy. To generalise, it would be good in the future to train on a rolling window and do cross-validation over multiple out-of-time test sets. It would also be extremely interesting to test this on other digital assets such as Ethereum, Solana, and more unknown (and volatile) assets such as FLOW. 

This work opens up many avenues for future research. For instance, the Twitter dataset could be more effective for predictions if it only includes tweets by influencers in the `crypto-Twitter sphere', such as Elon Musk, CEOs of cryptocurrency exchanges, and many more. In addition, we may finetune the finBERT model to better capture cryptocurrency-specific as well as Twitter-specific lingo. Finally, the resulting multimodal model's prediction threshold may be further finetuned with a more complex trading strategy, possibly including the model's class probability to size positions, to outperform the benchmark provided for our new dataset in this research.

\printcredits






\bibliographystyle{apalike}



\bibliography{paper}


\appendix
\section{Correlation analysis between features}

\begin{table}[h!]
\caption{The Pearson correlation values between the features in our dataset. }
\label{tab:corr_matrix_tab}   \setlength{\tabcolsep}{2pt}
\begin{adjustbox}{angle=90}
\scriptsize
\begin{tabular}{lcccccccccccccccccccc}  
\toprule                                      & High & Close &    &     &   &  & Adj  &     &    &   &   &    &    & upper & lower &     &  &     &    & ma \\
     & t+1 & t+1 & High   & Low    & Close  & Volume &  Close & ma7    & ma21   & 26ema  & 12ema  & MACD   & 20sd   & band & band & ema    & spread & eth    & gold   & indicator \\
                                      \midrule
High t+1  & 1.000             & 0.693              & 0.197  & -0.288 & -0.063 & 0.102  & -0.063    & 0.036  & 0.038  & 0.030  & 0.039  & -0.015 & 0.356  & 0.237       & -0.229      & -0.056 & 0.368  & -0.092 & -0.007 & 0.118       \\
Close t+1 & 0.693             & 1.000              & -0.017 & -0.058 & -0.056 & 0.013  & -0.056    & -0.039 & -0.032 & -0.034 & -0.037 & 0.023  & 0.051  & 0.008       & -0.065      & -0.060 & 0.033  & -0.092 & -0.012 & -0.041      \\
High               & 0.197             & -0.017             & 1.000  & 0.126  & 0.693  & 0.265  & 0.693     & 0.204  & 0.104  & 0.103  & 0.182  & 0.010  & 0.406  & 0.312       & -0.210      & 0.695  & 0.639  & 0.365  & 0.033  & 0.185       \\
Low                & -0.288            & -0.058             & 0.126  & 1.000  & 0.667  & -0.241 & 0.667     & -0.044 & -0.084 & -0.055 & -0.045 & 0.057  & -0.367 & -0.276      & 0.199       & 0.626  & -0.683 & 0.447  & 0.000  & -0.146      \\
Close              & -0.063            & -0.056             & 0.693  & 0.667  & 1.000  & 0.001  & 1.000     & 0.166  & 0.027  & 0.036  & 0.129  & 0.080  & 0.067  & 0.058       & -0.026      & 0.984  & -0.007 & 0.597  & 0.017  & 0.069       \\
Volume             & 0.102             & 0.013              & 0.265  & -0.241 & 0.001  & 1.000  & 0.001     & -0.038 & -0.033 & -0.033 & -0.037 & 0.021  & -0.032 & -0.042      & -0.005      & -0.006 & 0.382  & -0.087 & 0.028  & -0.123      \\
Adj Close          & -0.063            & -0.056             & 0.693  & 0.667  & 1.000  & 0.001  & 1.000     & 0.166  & 0.027  & 0.036  & 0.129  & 0.080  & 0.067  & 0.058       & -0.026      & 0.984  & -0.007 & 0.597  & 0.017  & 0.069       \\
ma7                & 0.036             & -0.039             & 0.204  & -0.044 & 0.166  & -0.038 & 0.166     & 1.000  & 0.728  & 0.704  & 0.888  & -0.356 & 0.132  & 0.579       & 0.524       & 0.314  & 0.185  & 0.072  & 0.040  & 0.683       \\
ma21               & 0.038             & -0.032             & 0.104  & -0.084 & 0.027  & -0.033 & 0.027     & 0.728  & 1.000  & 0.968  & 0.955  & -0.807 & 0.216  & 0.816       & 0.695       & 0.123  & 0.142  & -0.021 & 0.033  & 0.510       \\
26ema              & 0.030             & -0.034             & 0.103  & -0.055 & 0.036  & -0.033 & 0.036     & 0.704  & 0.968  & 1.000  & 0.931  & -0.901 & 0.208  & 0.789       & 0.673       & 0.130  & 0.118  & -0.016 & 0.030  & 0.489       \\
12ema              & 0.039             & -0.037             & 0.182  & -0.045 & 0.129  & -0.037 & 0.129     & 0.888  & 0.955  & 0.931  & 1.000  & -0.681 & 0.199  & 0.775       & 0.670       & 0.251  & 0.169  & 0.041  & 0.037  & 0.611       \\
MACD               & -0.015            & 0.023              & 0.010  & 0.057  & 0.080  & 0.021  & 0.080     & -0.356 & -0.807 & -0.901 & -0.681 & 1.000  & -0.182 & -0.662      & -0.556      & 0.038  & -0.037 & 0.081  & -0.015 & -0.254      \\
20sd               & 0.356             & 0.051              & 0.406  & -0.367 & 0.067  & -0.032 & 0.067     & 0.132  & 0.216  & 0.208  & 0.199  & -0.182 & 1.000  & 0.741       & -0.552      & 0.082  & 0.584  & 0.041  & 0.030  & 0.271       \\
upper band        & 0.237             & 0.008              & 0.312  & -0.276 & 0.058  & -0.042 & 0.058     & 0.579  & 0.816  & 0.789  & 0.775  & -0.662 & 0.741  & 1.000       & 0.151       & 0.133  & 0.444  & 0.009  & 0.041  & 0.511       \\
lower band        & -0.229            & -0.065             & -0.210 & 0.199  & -0.026 & -0.005 & -0.026    & 0.524  & 0.695  & 0.673  & 0.670  & -0.556 & -0.552 & 0.151       & 1.000       & 0.045  & -0.309 & -0.048 & 0.006  & 0.236       \\
ema                & -0.056            & -0.060             & 0.695  & 0.626  & 0.984  & -0.006 & 0.984     & 0.314  & 0.123  & 0.130  & 0.251  & 0.038  & 0.082  & 0.133       & 0.045       & 1.000  & 0.026  & 0.588  & 0.019  & 0.165       \\
spread             & 0.368             & 0.033              & 0.639  & -0.683 & -0.007 & 0.382  & -0.007    & 0.185  & 0.142  & 0.118  & 0.169  & -0.037 & 0.584  & 0.444       & -0.309      & 0.026  & 1.000  & -0.078 & 0.024  & 0.250       \\
eth                & -0.092            & -0.092             & 0.365  & 0.447  & 0.597  & -0.087 & 0.597     & 0.072  & -0.021 & -0.016 & 0.041  & 0.081  & 0.041  & 0.009       & -0.048      & 0.588  & -0.078 & 1.000  & 0.000  & 0.038       \\
gold               & -0.007            & -0.012             & 0.033  & 0.000  & 0.017  & 0.028  & 0.017     & 0.040  & 0.033  & 0.030  & 0.037  & -0.015 & 0.030  & 0.041       & 0.006       & 0.019  & 0.024  & 0.000  & 1.000  & 0.013       \\
ma indicator        & 0.118             & -0.041             & 0.185  & -0.146 & 0.069  & -0.123 & 0.069     & 0.683  & 0.510  & 0.489  & 0.611  & -0.254 & 0.271  & 0.511       & 0.236       & 0.165  & 0.250  & 0.038  & 0.013  & 1.000     \\
\bottomrule
\end{tabular}
\end{adjustbox}
\end{table}

\begin{table}[h!]
\caption{The p-values for the Pearson correlation between features in our dataset.}
\label{tab:corr_matrix_pvalue}   \setlength{\tabcolsep}{2pt}
\begin{adjustbox}{angle=90}
\scriptsize
\begin{tabular}{lcccccccccccccccccccc}  
\toprule                                      & High & Close &    &     &   &  & Adj  &     &    &   &   &    &    & upper & lower &     &  &     &    & ma \\
     & t+1 & t+1 & High   & Low    & Close  & Volume &  Close & ma7    & ma21   & 26ema  & 12ema  & MACD   & 20sd   & band & band & ema    & spread & eth    & gold   & indicator \\
                                      \midrule
High t+1  & 0.0000            & 0.0000             & 0.0000 & 0.0000 & 0.0024 & 0.0000 & 0.0024    & 0.0785 & 0.0638 & 0.1405 & 0.0616 & 0.4624 & 0.0000 & 0.0000      & 0.0000      & 0.0065 & 0.0000 & 0.0000 & 0.7284 & 0.0000      \\
Close t+1 & 0.0000            & 0.0000             & 0.4189 & 0.0049 & 0.0063 & 0.5161 & 0.0063    & 0.0615 & 0.1168 & 0.1033 & 0.0710 & 0.2613 & 0.0132 & 0.6968      & 0.0015      & 0.0039 & 0.1126 & 0.0000 & 0.5628 & 0.0491      \\
High               & 0.0000            & 0.4189             & 0.0000 & 0.0000 & 0.0000 & 0.0000 & 0.0000    & 0.0000 & 0.0000 & 0.0000 & 0.0000 & 0.6228 & 0.0000 & 0.0000      & 0.0000      & 0.0000 & 0.0000 & 0.0000 & 0.1098 & 0.0000      \\
Low                & 0.0000            & 0.0049             & 0.0000 & 0.0000 & 0.0000 & 0.0000 & 0.0000    & 0.0317 & 0.0000 & 0.0077 & 0.0286 & 0.0060 & 0.0000 & 0.0000      & 0.0000      & 0.0000 & 0.0000 & 0.0000 & 0.9926 & 0.0000      \\
Close              & 0.0024            & 0.0063             & 0.0000 & 0.0000 & 0.0000 & 0.9630 & 0.0000    & 0.0000 & 0.1886 & 0.0787 & 0.0000 & 0.0001 & 0.0012 & 0.0047      & 0.2052      & 0.0000 & 0.7342 & 0.0000 & 0.4118 & 0.0009      \\
Volume             & 0.0000            & 0.5161             & 0.0000 & 0.0000 & 0.9630 & 0.0000 & 0.9630    & 0.0680 & 0.1055 & 0.1152 & 0.0717 & 0.3076 & 0.1229 & 0.0426      & 0.8050      & 0.7791 & 0.0000 & 0.0000 & 0.1791 & 0.0000      \\
Adj Close          & 0.0024            & 0.0063             & 0.0000 & 0.0000 & 0.0000 & 0.9630 & 0.0000    & 0.0000 & 0.1886 & 0.0787 & 0.0000 & 0.0001 & 0.0012 & 0.0047      & 0.2052      & 0.0000 & 0.7342 & 0.0000 & 0.4118 & 0.0009      \\
ma7                & 0.0785            & 0.0615             & 0.0000 & 0.0317 & 0.0000 & 0.0680 & 0.0000    & 0.0000 & 0.0000 & 0.0000 & 0.0000 & 0.0000 & 0.0000 & 0.0000      & 0.0000      & 0.0000 & 0.0000 & 0.0005 & 0.0506 & 0.0000      \\
ma21               & 0.0638            & 0.1168             & 0.0000 & 0.0000 & 0.1886 & 0.1055 & 0.1886    & 0.0000 & 0.0000 & 0.0000 & 0.0000 & 0.0000 & 0.0000 & 0.0000      & 0.0000      & 0.0000 & 0.0000 & 0.3062 & 0.1088 & 0.0000      \\
26ema              & 0.1405            & 0.1033             & 0.0000 & 0.0077 & 0.0787 & 0.1152 & 0.0787    & 0.0000 & 0.0000 & 0.0000 & 0.0000 & 0.0000 & 0.0000 & 0.0000      & 0.0000      & 0.0000 & 0.0000 & 0.4309 & 0.1497 & 0.0000      \\
12ema              & 0.0616            & 0.0710             & 0.0000 & 0.0286 & 0.0000 & 0.0717 & 0.0000    & 0.0000 & 0.0000 & 0.0000 & 0.0000 & 0.0000 & 0.0000 & 0.0000      & 0.0000      & 0.0000 & 0.0000 & 0.0488 & 0.0715 & 0.0000      \\
MACD               & 0.4624            & 0.2613             & 0.6228 & 0.0060 & 0.0001 & 0.3076 & 0.0001    & 0.0000 & 0.0000 & 0.0000 & 0.0000 & 0.0000 & 0.0000 & 0.0000      & 0.0000      & 0.0645 & 0.0770 & 0.0001 & 0.4542 & 0.0000      \\
20sd               & 0.0000            & 0.0132             & 0.0000 & 0.0000 & 0.0012 & 0.1229 & 0.0012    & 0.0000 & 0.0000 & 0.0000 & 0.0000 & 0.0000 & 0.0000 & 0.0000      & 0.0000      & 0.0001 & 0.0000 & 0.0499 & 0.1402 & 0.0000      \\
upper band        & 0.0000            & 0.6968             & 0.0000 & 0.0000 & 0.0047 & 0.0426 & 0.0047    & 0.0000 & 0.0000 & 0.0000 & 0.0000 & 0.0000 & 0.0000 & 0.0000      & 0.0000      & 0.0000 & 0.0000 & 0.6470 & 0.0480 & 0.0000      \\
lower band        & 0.0000            & 0.0015             & 0.0000 & 0.0000 & 0.2052 & 0.8050 & 0.2052    & 0.0000 & 0.0000 & 0.0000 & 0.0000 & 0.0000 & 0.0000 & 0.0000      & 0.0000      & 0.0307 & 0.0000 & 0.0205 & 0.7765 & 0.0000      \\
ema                & 0.0065            & 0.0039             & 0.0000 & 0.0000 & 0.0000 & 0.7791 & 0.0000    & 0.0000 & 0.0000 & 0.0000 & 0.0000 & 0.0645 & 0.0001 & 0.0000      & 0.0307      & 0.0000 & 0.2082 & 0.0000 & 0.3575 & 0.0000      \\
spread             & 0.0000            & 0.1126             & 0.0000 & 0.0000 & 0.7342 & 0.0000 & 0.7342    & 0.0000 & 0.0000 & 0.0000 & 0.0000 & 0.0770 & 0.0000 & 0.0000      & 0.0000      & 0.2082 & 0.0000 & 0.0002 & 0.2419 & 0.0000      \\
eth                & 0.0000            & 0.0000             & 0.0000 & 0.0000 & 0.0000 & 0.0000 & 0.0000    & 0.0005 & 0.3062 & 0.4309 & 0.0488 & 0.0001 & 0.0499 & 0.6470      & 0.0205      & 0.0000 & 0.0002 & 0.0000 & 0.9863 & 0.0641      \\
gold               & 0.7284            & 0.5628             & 0.1098 & 0.9926 & 0.4118 & 0.1791 & 0.4118    & 0.0506 & 0.1088 & 0.1497 & 0.0715 & 0.4542 & 0.1402 & 0.0480      & 0.7765      & 0.3575 & 0.2419 & 0.9863 & 0.0000 & 0.5179      \\
ma indicator        & 0.0000            & 0.0491             & 0.0000 & 0.0000 & 0.0009 & 0.0000 & 0.0009    & 0.0000 & 0.0000 & 0.0000 & 0.0000 & 0.0000 & 0.0000 & 0.0000      & 0.0000      & 0.0000 & 0.0000 & 0.0641 & 0.5179 & 0.0000     \\
\bottomrule
\end{tabular}
\end{adjustbox}
\end{table}

\FloatBarrier

\section{Optimized hyperparameters}
\begin{table}[h!] \scriptsize
\caption{Optimal hyperparameters based on the validation set, and that are used for the results obtained in Section~\ref{sec:experiments}. }
\label{tablehyperpara}
\label{tab:hyperpara} \setlength{\tabcolsep}{2pt}
    \centering 
    \begin{tabular}{l|c|c|c|c }
        \toprule
        Models & Task Up 5\% & Task Up 2\% & Task Down 5\% & Task Down 2\% \\
        \midrule
        TA SVM  & $RBF, C = 50, gamma = 0.5$ & $RBF, C = 500, gamma = 0.1$ & $RBF, C = 1000, gamma = 0.1$ & $RBF, C = 1000, gamma = 0.1$\\ 
        Twitter CNN (P) & $\alpha = 0.12, \gamma = 1$ & CE loss & $\alpha = 0.12, \gamma = 1$  & CE loss\\
        Twitter CNN (S) & $\alpha = 0.12, \gamma = 1, L2 weight = 0.0005$ & CE loss $ L2 weight = 0.001$ & $\alpha = 0.12, \gamma = 1, L2 weight = 0.0005$ & CE loss $ L2 weight = 0.001$\\
        Fusion model ({P}) & $polynomial,C = 500, gamma = 50$ & Logistic regression & Logistic regression & $polynomial, C = 10, gamma = 10$\\
        Fusion model (S) & $polynomial,C = 30, gamma = 50$ & Logistic regression & Logistic regression& $polynomial, C = 10, gamma = 50$\\
        \bottomrule
    \end{tabular}
\end{table}



\end{document}